\documentclass[amssymb,pra,aps,epsfig]{revtex4}
\newcommand{\be}{\begin{equation}}
\newcommand{\ee}{\end{equation}}
\newcommand{\br}{\begin{eqnarray}}
\newcommand{\er}{\end{eqnarray}}
\newcommand{\nn}{\nonumber}
\newcommand{\bd}{\begin{displaymath}}
\newcommand{\ed}{\end{displaymath}}

\newcommand{\bib}{\bibitem}
\newcommand{\bfig}{\begin{figure}}
\newcommand{\efig}{\end{figure}}
\usepackage{graphicx}
\def\alf{\alpha}
\def\th{\theta}
\def\lam{\lambda}

\def\eps{\epsilon}
\def\rpar{\right)}
\def\lpar{\left(}
\def\rbk{\right]}
\def\lbk{\left[}
\def\rbr{\right\}}
\def\lbr{\left\{}
\def\lb{\label}
\def\upa{\uparrow}
\def\dwa{\downarrow}
\def\im{{\rm i}}
\def\tr{\mbox{${\rm Tr}$}}
\def\norm{\mbox{${\rm N}$}}
\def\ro{\mbox{\boldmath $\rho$}}
\def\sig{\mbox{\boldmath $\sigma$}}

\def\bu{\mbox{\boldmath ${\cal U}$}}
\def\bht{\mbox{\boldmath ${\cal H}$}}

\def\ima{\mbox{${\rm Im}$}}
\def\re{\mbox{${\rm Re}$}}
\def\rg{\rangle}
\def\lg{\langle}
\def\half{\frac{1}{2}}
\begin{document}
\draft
\title{Engineering superpositions of displaced number states of a trapped ion}
\author{Marcelo A. Marchiolli\footnote{Corresponding author: Avenida General Os\'{o}rio 414-centro, 14870-100 Jaboticabal,
        SP, Brazil.}}
\address{Instituto de F\'{\i}sica de S\~{a}o Carlos, Universidade de S\~{a}o Paulo, \\
         Caixa Postal 369, 13560-970 S\~{a}o Carlos, SP, Brazil, \\
         E-mail address: marcelo$\_$march@bol.com.br}
\author{Wagner Duarte Jos\'{e}}
\address{Universidade Estadual de Santa Cruz, Departamento de Ci\^{e}ncias Exatas e Tecnol\'{o}gicas, \\
         Rodovia Ilh\'{e}us/Itabuna Km 16, 45650-000 Ilh\'{e}us, Bahia, Brazil, \\
         E-mail address: wjose@uesc.br}
\date{\today}
%
\begin{abstract}
\vspace*{0.1mm}
\begin{center}
\rule[0.1in]{142mm}{0.4mm}
\end{center}
We present a protocol that permits the generation of a subtle superposition with $2^{\ell + 1}$ displaced number states on
a circle in phase space as target state for the center-of-mass motion of a trapped ion. Through a sequence of $\ell$ cycles
involving the application of laser pulses and no-fluorescence measurements, explicit expressions for the total duration of
laser pulses employed in the sequence and probability of getting the ion in the upper electronic state during the $\ell$
cycles are obtained and analyzed in detail. Furthermore, assuming that the effective relaxation process of a trapped ion 
can be described in the framework of the standard master equation for the damped harmonic oscillator, we investigate the 
degradation of the quantum interference effects inherent to superpositions via Wigner function. \\
\vspace*{0.1mm}
\begin{center}
\rule[0.1in]{142mm}{0.4mm}
\end{center}
\end{abstract}
\maketitle
\section{Introduction}

In quantum mechanics, the nomenclature `nonclassical states' has been employed by theoretical physics for approxi\-mately
seventy five years in order to designate the states of quantum systems whose statistical properties present genuine quantum 
effects (e.g., photon antibunching, sub-Poissonian statistics, squeezing, and quantum interference effects inherent to 
superposition states) without having analogous effects in classical mechanics \cite{r1}. During this period, different
experimental techniques have been developed for generating and detecting both trapped and travelling nonclassical states, 
and more recently the nonclassical electronic and vibrational states of trapped ions. For instance, the density matrices and
Wigner functions associated to the Fock states, thermal states, coherent states, squeezed vacuum states, and Schr\"{o}dinger
cat states (entangled position and spin superposition states), were reconstructed in laboratory through a beautiful
experiment involving the quantum states of motion (motional states) of a harmonically bound $^{9}{\mbox {\rm Be}}^{+}$ ion 
\cite{r2,r3,r4,r5}. After this experiment, a considerable number of papers dedicated to generation and detection of 
nonclassical motional states have appeared in the literature \cite{r6,r7,r8,r9,r10,r11,r12,r13,r14,r15,r16,r17,r18,r19}. In 
particular, Matos Filho and Vogel \cite{r6} have considered a class of nonlinear coherent states (NCS) which exhibits 
interesting nonclassical features such as strong squeezing and self-splitting with pronounced quantum interference effects, 
and showed that they may appear as stationary states (also recognized by the authors as dark states) of the center-of-mass 
(CM) motion of a trapped and bichromatically laser-driven ion far from the Lamb-Dicke regime. Man'ko et al. \cite{r12} have 
extended the results obtained by Matos Filho and Vogel for NCS on a circle, where the influence of nonlinear effects on the 
Wigner functions was discussed in detail. Furthermore, Kis et al. \cite{r14} have introduced a method by which any pure 
state of the quantum harmonic oscillator can be represented in a limiting sense as a NCS, and showed through a physical 
example how to prepare a highly excited Fock state in an ion trap based on the concept of NCS. On the other hand, Moya-Cessa 
et al. \cite{r10} have shown how an arbitrary superposition of coherent states can be created on a line in phase space for 
the motion of a single trapped ion. Pursuing this line, Duarte Jos\'{e} and Mizrahi \cite{r11} have proposed three schemes 
to engineer circular states (superposition of $N$ coherent states on a circle in phase space) for the CM motion of a trapped 
ion, where the total duration of laser pulses and the probability of getting the ion in the upper electronic state were 
determined for each process. In addition, the authors also have shown how the interference effects between the components 
of the subtle superposition can produce the Fock states.

Recently, many authors have investigated different sources of decoherence in experiments involving trapped ions and 
predicted interesting results which permit us to give reasonable explanations on the phenomenological decay rate of Rabi
oscillations \cite{r20,r21,r22,r23}. In this sense, Schneider and Milburn \cite{r20} have considered as decoherence source
the intensity and phase fluctuations in the exciting laser pulses, and showed that a simple master equation description
can be obtained since the stochastic processes involved are white noise processes. Serra et al. \cite{r21} have examined,
through the analogy with the physics of surface electrons in liquid helium, the mechanism of damping and heating of trapped
ions associated with the polarization of the residual background gas induced by the oscillating ions themselves. In 
particular, the authors have demonstrated that the decay of Rabi oscillations observed in experiments on 
$^{9}{\mbox {\rm Be}}^{+}$ can be attributed to the polarization phenomena. Budini et al. \cite{r22,r23} have assumed that 
the origin of decoherence of the nonclassical motional states is due to the coupling of the vibronic modes with classical 
fields and to the finite lifetime of the electronic levels, and showed that these interactions lead to a dispersive-like 
decoherence dynamics. On the other hand, Poyatos et al. \cite{r24} have shown how to design different couplings (due to the
absorption of a laser photon and subsequent spontaneous emission) between a single ion trapped in a harmonic potential and
environment. In this scheme, the variation of the laser frequencies and intensities allows one to `engineer' the coupling
and select the master equation which describes the motion of the ion. Turchette et al. \cite{r25} also have presented
results from an experimental study of the decoherence and decay of motional states due the interaction with several types
of engineered reservoirs. Now, independently of the damping mechanisms to be considered, the degradation of the quantum 
interference effects in superpositions of motional states is always verified.

One of the most important applications using trapped ions was established by Cirac and Zoller \cite{r26} in the context of
quantum computation, where the authors have shown that a set of $N$ cold ions interacting with laser light and moving in a
linear trap provide a realistic physical system to implement a quantum computer. The main features of this proposal are 
that (i) decoherence can be made negligible during the computation process, (ii) the implementation of $n$-bit quantum gates
between any set of ions is relatively straightforward, and (iii) the quantum bit readout can be performed with efficiency
approximately equal to one (a quantum bit or qubit refers to a two-state system characterized by $\{ | 0 \rg, | 1 \rg \}$).
However, the experimental realization of a quantum computer requires isolated quantum systems acting as the qubits, and the 
presence of controlled unitary interactions between the qubits allowing the construction of the controlled-NOT gate 
(basically, a controlled-NOT is defined by the operation $| \eps_{1} \rg | \eps_{2} \rg \rightarrow | \eps_{1} \rg | 
\eps_{1} \oplus \eps_{2} \rg$ with $\oplus$ denoting addition modulo 2, and $\eps_{1,2} = 0,1$). Thus, if the qubits are
not sufficiently isolated from the environment, different mechanisms of decoherence can destroy the quantum interferences
that make the computation. The first experimental implementation of a fundamental quantum logic gate that operates on 
prepared quantum states in experiments involving trapped ions was realized by Monroe et al. \cite{r27}. Following the
scheme proposed in \cite{r26}, the authors have demonstrated a controlled-NOT gate on a pair of qubits which illustrates the 
basic operations necessary, and the problems associated, in the construction process of a large scale quantum computer 
(in this experiment, the switching speed of the controlled-NOT gate is approximately 20 kHz and the decoherence rate is of 
a few kHz). After the original ion-trap proposal of Cirac and Zoller, a number of modifications and extensions to their idea 
have appeared in the literature (e.g., see Refs. \cite{r28,r29,r30,r31,r32,r33}). In summary, the investigation of noise 
sources in such promising quantum systems turns out to be a crucial step toward the implementation of a quantum logic 
processor, and consequently, of a quantum computer.

According to Kis et al. \cite{r14}: ``Nonclassical states of the electromagnetic field and the atomic center-of-mass motion
have played an important role in recent years, due to their relation with fundamental problems in quantum mechanics and to
the many possible applications, ranging from high-resolution spectroscopy to low-noise communication and quantum 
computation. However, the generation of these states is usually a demanding experimental challenge." In the present 
contribution, we propose a systematic scheme which permits us to engineer superpositions of displaced number states on a 
circle in phase space as target states for the CM motion of a trapped ion. These superpositions were studied by Marchiolli 
et al. \cite{r34}, where the authors have shown that (i) the interference effects among the state components present an 
analogy with diffraction patterns arising in an $N$ slit Young-type experiment, and (ii) the interference and correlation 
effects are connected with the nondiagonal term of the quasiprobability distributions. In general, the superpositions of 
$N$ displaced number states on a circle in phase space can be defined as follows:
\be
\lb{e1}
| \Psi_{n}^{(N)} (\beta) \rg \equiv \norm_{n}^{(N)} \sum_{r=1}^{N} {\bf D}(\beta_{r}) | n \rg = \norm_{n}^{(N)} 
\sum_{r=1}^{N} | n, \beta_{r} \rg \qquad (n \in \mathbb{N}) \; ,
\ee
where ${\bf D}(\beta_{r}) = \exp \lpar \beta_{r} {\bf a}^{\dagger} - \beta_{r}^{\ast} {\bf a} \rpar$ is the displacement
operator with $\beta_{r} = |\beta| e^{\im \th_{r}}$ and $\th_{r+1} - \th_{r} = 2 \phi$ for $\phi \in [0,2\pi]$,
\bd
\norm_{n}^{(N)} = \lbr N + 2 \sum_{r=1}^{N-1} r \; e^{- 2 |\beta|^{2} \sin^{2} [(N-r) \phi]} \cos \lpar |\beta|^{2} \sin
[2(N-r) \phi] \rpar L_{n} \lpar 4 |\beta|^{2} \sin^{2} [(N-r) \phi] \rpar \rbr^{-1/2}
\ed
is the normalization constant, and $L_{n}(z)$ is the Laguerre polynomial. An additional property of these superpositions was
established in \cite{r11} for $n=0$ and $1 \ll ( {\rm e} | \beta |^{2} /N )^{N} \ll 4^{N}$, where the interference effects
approximately produce a particular Fock state; while for $( {\rm e} | \beta |^{2} /N )^{N} \ll 1$, an almost vacuum state
is reached. To engineer (\ref{e1}) we initially prepare the trapped-ion state in $| \Phi (0) \rg = | n \rg \otimes | \upa 
\rg$ by means of the experimental techniques described in Refs. \cite{r2,r3,r4,r5} (this procedure characterizes the first 
step of our scheme). With the help of the method established in \cite{r35} and used by Moya-Cessa et al. \cite{r10} for 
obtaining an arbitrary superposition of coherent states, the second step consists in the generation of superpositions of 
two displaced number states on a line. Now, considering the proposal of Duarte Jos\'{e} and Mizrahi \cite{r11} for 
engineering circular states and adopting the motional state reached in the previous procedure as the initial motional state 
for this last step, we obtain, after a sequence of $\ell$ cycles involving the application of laser pulses and 
no-fluorescence measurements, the state (\ref{e1}) in phase space. Furthermore, the total duration $T$ of laser pulses 
employed in the sequence and the probability $P_{\upa}(T)$ of getting the ion in the upper electronic state after $\ell$ 
cycles are explicitly calculated. We also verify that the quantum interference effects between the $N = 2^{\ell + 1}$ 
components of the motional state obtained in the third step decrease the values of $P_{\upa}(T)$ when superpositions with 
$N \gg 1$ are regarded. Finally, assuming that the effective relaxation process of trapped ions can be described by the 
standard master equation for the damped harmonic oscillator \cite{r16,r25}, we evaluate the time evolution of the Wigner 
function associated to $| \Psi_{n}^{(N)} (\beta) \rg$. Following, this function is factorized into diagonal and nondiagonal 
terms which permits us to investigate, for example, the degradation of the undermentioned quantum interference effects 
through a quantitative measure of coherence introduced in \cite{r36} that characterizes the decoherence process of the 
nondiagonal elements of a density operator in the Fock-state basis. 

This paper is organized as follows. In Section II we adopt the method proposed by Wallentowitz and Vogel \cite{r35} in
order to obtain a unitary time-evolution operator which permits one to produce superpositions of two displaced number 
states on a line. To engineer (\ref{e1}) we consider in Section III the proposal of Duarte Jos\'{e} and Mizrahi \cite{r11} 
for obtaining circular states, and also determine the total duration of laser pulses and the probability of getting the ion 
in the upper electronic state during the construction process. In Section IV we employ the Weyl-Wigner formalism to 
investigate the degradation of the quantum interference effects between the $2^{\ell + 1}$ components of the motional state 
(\ref{e1}) through an effective relaxation process of the trapped ion. Section V contains our summary and conclusions. 
Finally, Appendix A describes the calculational details on the measure of coherence used in Section IV.

\section{Engineering superpositions of two displaced number states on a line}

Let us consider a weak electronic transition of an ion which is bichromatically irradiated by two laser fields detuned to
the first lower and first upper vibrational sidebands of the transition, respectively, with equal intensities. In the
resolved-sideband and Lamb-Dicke regimes, the interaction Hamiltonian for the laser-assisted vibronic coupling can be
written, in the interaction picture, as \cite{r35}
\be
\lb{e2}
{\bf H}_{{\rm int}} = \sqrt{2} \; \Omega \lpar \sig_{-} \; e^{\im \varphi} + \sig_{+} \; e^{- \im \varphi} \rpar 
{\bf X}_{\th} \; ,
\ee
where $\Omega = \eta \lam$ is the effective Rabi frequency on the first vibrational sideband with coupling constant $\lam$
and Lamb-Dicke parameter $\eta$. The electronic flip operators $\sig_{\pm}$ and $\sig_{z}$ describe the electronic 
transitions $| \dwa \rg \rightleftharpoons | \upa \rg$ and satisfy the commutation relations $\lbk \sig_{+},\sig_{-} \rbk = 
\sig_{z}$ and $\lbk \sig_{z},\sig_{\pm} \rbk = \pm 2 \sig_{\pm}$. The phase-rotated quadrature operator
\be
\lb{e3}
{\bf X}_{\th} \equiv \frac{{\bf a} \; e^{\im \th} + {\bf a}^{\dagger} \; e^{- \im \th}}{\sqrt{2}} = {\bf Q} \cos \th -
{\bf P} \sin \th \qquad (- \pi \leq \th \leq \pi)
\ee
represents the generalized CM position of the ion, being ${\bf a}^{\dagger}$ $( {\bf a} )$ the creation (annihilation) 
operator of vibrational quanta. Here, the dimensionless quadrature operators ${\bf Q}$ (positionlike) and ${\bf P}$ 
(momentumlike) obey the Weyl-Heisenberg commutation relation $\lbk {\bf Q},{\bf P} \rbk = \im {\bf 1}$ (for simplicity, we 
will fix $\hbar = 1$ throughout this paper). Furthermore, the phases $\varphi = \half (\varphi_{{\rm b}} + 
\varphi_{{\rm r}})$ and $\th = \half (\varphi_{{\rm b}} - \varphi_{{\rm r}})$ contain the phases $\varphi_{{\rm b}}$ and 
$\varphi_{{\rm r}}$ of the lasers detuned to the blue $({\rm b})$ and red $({\rm r})$ sides of the electronic transition. 
In particular, when $\varphi_{{\rm r}} = \varphi_{{\rm b}}$ or $\varphi_{{\rm r}} = \varphi_{{\rm b}} + \pi$, we obtain the 
operators ${\bf X}_{0} = {\bf Q}$ and ${\bf X}_{- \pi/2} = {\bf P}$.

Using the interaction Hamiltonian (\ref{e2}), we can express the unitary time-evolution operator $\bu(t) = \exp \lpar - 
\im t {\bf H}_{{\tt int}} \rpar$ in a compact form as follows \cite{r10}:
\be
\lb{e4}
\bu(t) = \cos \lpar \sqrt{2} \Omega t {\bf X}_{\th} \rpar - \im \lpar \sig_{-} \; e^{\im \varphi} + \sig_{+} \;
e^{- \im \varphi} \rpar \sin \lpar \sqrt{2} \Omega t {\bf X}_{\th} \rpar \; .
\ee
This result allows one to determine the density operator $\ro(t) = \bu(t) \ro(0) \bu^{\dagger}(t)$ with $\ro(0) = \ro_{v}
(0) \otimes | \upa \rg \lg \upa |$ (being $\ro_{v}(0)$ the density operator for the CM motional state at time $t=0$), and
to prepare a superposition of displaced number states on a line in phase space. In fact, we are interested in generating 
superpositions of two displaced number states on a line using the present approach. For this purpose, the evolution 
operator (\ref{e4}) is applied on the state $| \Phi (0) \rg = | n \rg \otimes | \upa \rg$, giving the following result:
\be
\lb{e5}
| \Phi (\beta) \rg = \half \lpar | n, \beta \rg + | n, -\beta \rg \rpar \otimes | \upa \rg - \half \; e^{\im \varphi} \lpar
| n, \beta \rg - | n, -\beta \rg \rpar \otimes | \dwa \rg \; ,
\ee
where $| n, \pm \beta \rg = {\bf D} (\pm \beta) | n \rg$ correspond to the displaced number states \cite{r37} and whose
statistical properties were studied in detail by de Oliveira et al. \cite{r38}, ${\bf D}(\beta) = \exp \lpar \beta 
{\bf a}^{\dagger} - \beta^{\ast} {\bf a} \rpar$ is the displacement operator, and $\beta = \im \Omega t e^{-\im \th}$.
The procedure of measurement of the motional state was established in \cite{r2,r3,r4,r5}, and it consists of collecting the 
emitted resonance fluorescence signal from the transition $| d \rg \leftrightarrow | \dwa \rg$ (being $| d \rg$ an 
auxiliary electronic state with width $\Gamma$ on the order of $\Gamma / 2\pi \approx 20$ MHz) by means of a laser strongly 
coupled to the electronic ground state during a specific period of time $\tau$. Following, we consider only those events 
where no fluorescence have been observed, since any spontaneously emitted photon will disturb the motional quantum state 
via recoil effects. At this point, it is important mentioning that the efficiency in collecting the fluorescence of the
trapped ion is of order of $10^{-4}$ (i.e., about $10^{4}$ photons have to be scattered by the cycling electronic 
transition to be detected). Thus, the time needed to detect the electronic state of the ion is approximately equal to 
$\tau_{{\rm d}} \approx 200 \mu{\rm s}$ \cite{r39,r40}. Now, if one considers the measurement time $(\tau_{{\rm d}})$ in
the evaluation of the total time $(\tau_{{\rm t}})$ necessary to prepare the target state, one obtains $\tau_{{\rm t}} 
\approx \tau + 200 \mu{\rm s}$. Consequently, the quantum state (\ref{e5}) is projected onto the excited state $| \upa \rg$ 
and the resulting conditioned vibronic quantum state reads
\be
\lb{e6}
| \widetilde{\Phi} (\beta) \rg = \frac{{\cal N}_{n}^{(2)}}{\sqrt{2}} \lpar | n, \beta \rg + | n, -\beta \rg \rpar \otimes
| \upa \rg \; ,
\ee
being ${\cal N}_{n}^{(2)}$ the normalization factor. In particular, the probability $P_{\upa}(\beta)$ for the occurrence of
the no-fluorescence event is connected with the normalization factor through the relation $| {\cal N}_{n}^{(2)} |^{2}
P_{\upa}(\beta) = 1/2$, and its maximum and minimum points depend on the excitation degree $n$ of the motional state
described by Eq. (\ref{e6}). Figure 1 shows the plot of $P_{\upa}(\beta)$ versus $| \beta |$ for $n=0$ (dot-dashed line), 
1 (dashed line) and 2 (solid line), where we observe that the maximum of this function reaches $P_{\upa} \approx 0.68$ when
$n=2$ and $| \beta | \approx 1.27$ (for instance, if one considers $\eta \approx 0.1$ and $\lam / 2\pi \approx 1$ MHz, this
value of $| \beta |$ corresponds to $\tau_{{\rm t}} \approx 202$ $\mu$s which is greater than $2\pi / \Gamma \approx 0.05$ 
$\mu$s).
\bfig[!t]
\centering
\begin{minipage}[b]{0.45\linewidth}
\includegraphics[width=\linewidth]{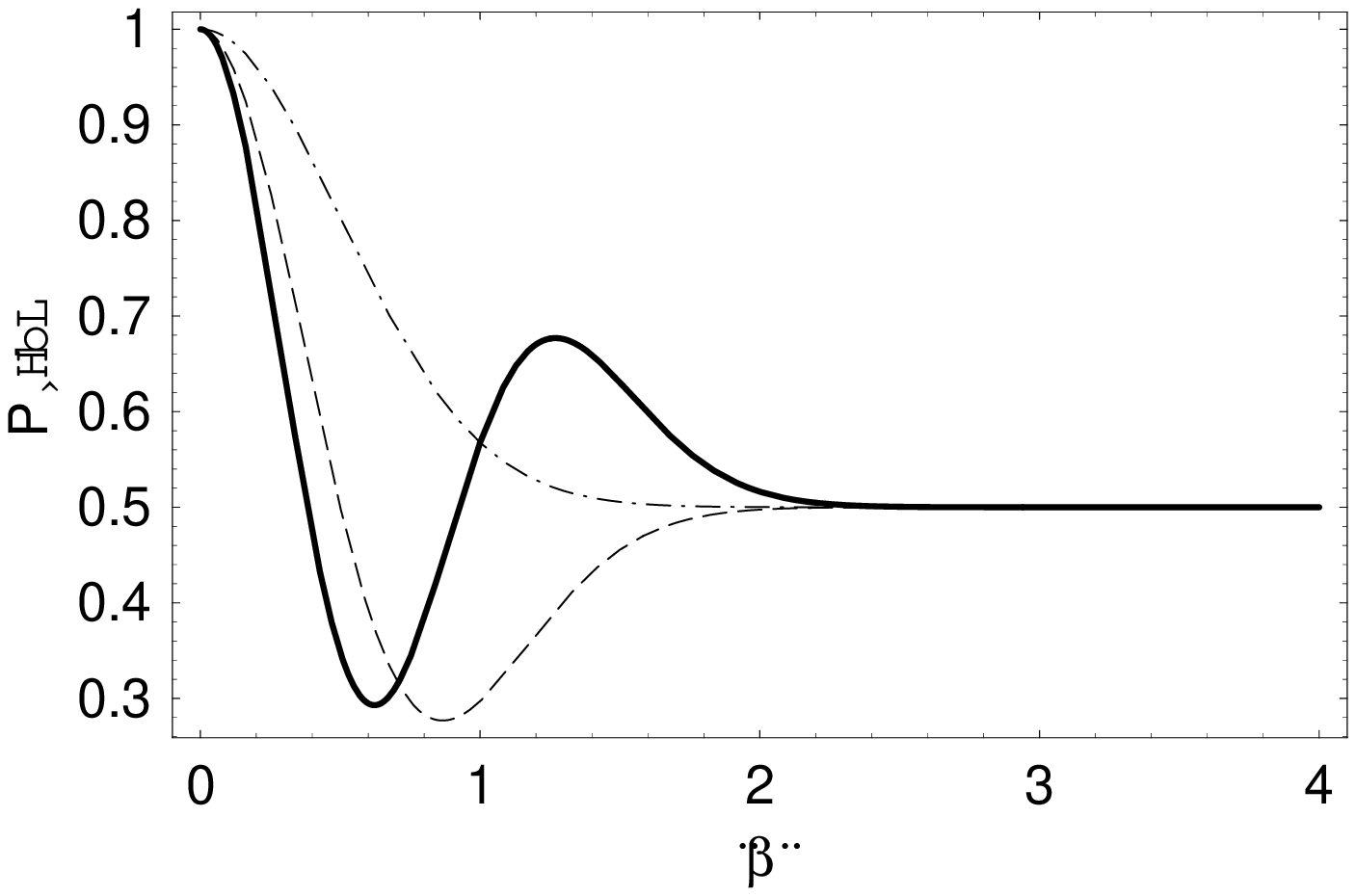}
\end{minipage}
\caption{Plot of $P_{\upa}(\beta) = \half [1 + \exp (- 2 | \beta |^{2}) L_{n} (4 | \beta |^{2})]$ versus $| \beta | \in
[0,4]$ and different values of the excitation degree $n$, where the dot-dashed, dashed and solid lines correspond to
$n=0,1$ and $2$, respectively.}
\efig

\section{Engineering superpositions of displaced number states on a circle}

In this third step, we adopt the procedure established by Duarte Jos\'{e} et al. \cite{r11,r16} which is based on a
Kerr-type interaction obtained through the interaction between the trapped ion and one pair of laser beams tuned in
resonance with the electronic transition frequency. In the Lamb-Dicke regime (for more details, see Refs. 
\cite{r41,r42,r43}), the carrier Hamiltonian of this system can be approximated as follows \cite{r11}:
\be
\lb{e7}
\bht = \Lambda \sig_{x} \exp \lpar - \frac{\kappa^{2}}{2} \rpar \lbk {\bf 1} - \kappa^{2} {\bf n} + \frac{\kappa^{4}}{4} 
({\bf n}^{2} - {\bf n}) - \frac{\kappa^{6}}{36} ({\bf n}^{3} - 3 {\bf n}^{2} + 2 {\bf n}) + \cdots \rbk \approx \Lambda 
\sig_{x} \lpar {\bf 1} - \kappa^{2} {\bf n} \rpar \; ,
\ee
where $\Lambda$ is the effective Rabi frequency, $\kappa$ is the Lamb-Dicke parameter, ${\bf n} = {\bf a}^{\dagger}
{\bf a}$ is the phonon-number operator, and ${\bf 1}$ is the identity operator. It is important mentioning that the
validity of the Hamiltonian $\bht$ essentially depends on the condition $(\kappa^{2}/4) \lg {\bf a}^{\dagger 2} {\bf a}^{2} 
\rg \ll \lg {\bf a}^{\dagger} {\bf a} \rg$, i.e., this condition must be satisfied in order to guarantee the validity of 
the approximation employed in (\ref{e7}). In addition, we also consider the vibronic quantum state (\ref{e6}) as an initial
state for the third step of the protocol.

To construct superpositions of displaced number states on a circle we have used the sequence outlined in Ref. \cite{r11}
which consists of $\ell$ cycles involving the application of laser pulses and no-fluorescence measurements. In fact, each
cycle consists in the application of one laser pulse with specific duration $t_{k}$ (in order to generate the required
superposition) followed by one no-fluorescence measurement (this event assures no-recoil effects of the vibrational motion
of a trapped ion, and maximizes the probability to realize successfully the target state). Now, if in a particular cycle
of measurements a fluorescence emission is detected, the sequence must be stopped and repeated again. Thus, after $\ell$
cycles of successfull measurements, the resulting conditioned vibronic quantum state becomes
\be
\lb{e8}
| \widetilde{\Phi} (t_{1} + \cdots + t_{\ell}) \rg = \sqrt{2} \; {\cal M}_{n}^{(\ell)} \lbk \; \prod_{k=1}^{\ell} \lg \upa 
| {\bf U} (t_{k}) | \upa \rg \rbk \lpar |n,\beta \rg + |n,-\beta \rg \rpar \otimes | \upa \rg \; ,
\ee
where
\be
\lb{e9}
{\bf U}(t_{k}) = \exp \lbk - \im t_{k} \lpar \Lambda {\bf 1} - \bar{\Lambda} {\bf n} \rpar \sig_{x} \rbk 
\qquad ( \bar{\Lambda} \equiv \kappa^{2} \Lambda)
\ee
is the unitary time-evolution operator at time $t_{k}$ associated to the carrier Hamiltonian $\bht$. To engineer the ion
CM motional state as Eq. (\ref{e1}), we need to adjust the phases in (\ref{e8}) of the displaced number states putting them 
evenly distributed around the circle: this fact is only possible when the duration of the $k$th evolution pulse is given by 
$t_{k} = \pi / \lpar 2^{k+1} \bar{\Lambda} \rpar$. Furthermore, we have chosen conveniently the Lamb-Dicke parameter as 
$\kappa^{2} = \lpar n + 2^{\ell + 2} \rpar^{-1}$, which permits us to write the superposition (\ref{e8}) in the simplified 
form
\be
\lb{e10}
| \widetilde{\Phi} (t_{1} + \cdots + t_{\ell}) \rg = | \Psi_{n}^{(2^{\ell + 1})} (\beta) \rg \otimes | \upa \rg
\ee
since $\phi = \pi/2^{\ell + 1}$, $\th_{r} = 2 \pi r/2^{\ell + 1}$ with $r = 1, \ldots, 2^{\ell + 1}$, and $N=2^{\ell + 1}$ 
are fixed a priori. The final adjustment of phases involved in this process is reached when the phases $\varphi_{{\rm r}}$ 
and $\varphi_{{\rm b}}$ of the lasers described in the previous section satisfy the relation $\varphi_{{\rm r}} = 
\varphi_{{\rm b}} + \pi/4$. Consequently, the normalization constants ${\cal M}_{n}^{(\ell)}$ and 
$\norm_{n}^{(2^{\ell + 1})}$ can be connected by means of the equality ${\cal M}_{n}^{(\ell)} = 2^{(\ell + 1)/2} 
\norm_{n}^{(2^{\ell + 1})}$.

The total duration $T = t_{1} + \cdots + t_{\ell}$ of laser pulses employed in the sequence, the total time 
${\rm T}_{{\rm t}}$ necessary to prepare the target states (\ref{e10}), and the probability $P_{\upa}(T)$ of getting the 
ion in the upper electronic state during the $\ell$ cycles, i.e.,
\br
\lb{e11}
T &=& \frac{\pi}{2 \Lambda} \lpar n + 2^{\ell + 2} \rpar \lpar 1 - 2^{-\ell} \rpar \; , \\
\lb{e12}
{\rm T}_{{\rm t}} &=& T + \ell \tau_{{\rm d}} + \tau_{{\rm t}} \; ,
\er
and
\be
\lb{e13}
P_{\upa}(T) = \frac{1}{2^{\ell + 1}} + \frac{1}{2^{2 \ell + 1}} \sum_{r=1}^{2^{\ell + 1}-1} r \; \exp \lbk - 2 |\beta|^{2} 
\sin^{2} \lpar \frac{\pi r}{2^{\ell + 1}} \rpar \rbk \cos \lbk | \beta |^{2} \sin \lpar \frac{2 \pi r}{2^{\ell + 1}} \rpar 
\rbk L_{n} \lbk 4 | \beta |^{2} \sin^{2} \lpar \frac{\pi r}{2^{\ell + 1}} \rpar \rbk \; ,
\ee
permit us to characterize completely the construction process. It is important mentioning that $\norm_{n}^{(2^{\ell + 1})}$ 
and $P_{\upa}(T)$ are connected through the relation $| \norm_{n}^{(2^{\ell + 1})} |^{2} P_{\upa}(T) = 2^{- 2(\ell + 1)}$. 
As an application of the results obtained until the present moment we consider the engineering of superpositions with $N=4$ 
and $8$ displaced number states, which correspond to different sequences involving $\ell = 1$ and $2$ cycles each one, and 
analyze their respective success through the probability (\ref{e13}). Figure 2 shows the plot of $P_{\upa}(T)$ versus 
$| \beta |$ for (a) $\ell = 1$ and (b) $\ell = 2$, with $n=0$ (dot-dashed line), $1$ (dashed line) and $2$ (solid line) 
fixed in both situations. In Fig. 2(a) the maximum of this function reaches $P_{\upa} \approx 0.37$ $(0.34)$ when $n=1$ 
$(2)$ and $| \beta | \approx 1.65$ $(1.28)$, with $\kappa \approx 0.33$ $(0.32)$ and ${\rm T}_{{\rm t}} \approx 401.93$ 
$\mu$s (401.85 $\mu$s) if one considers $2 \pi / \Lambda \approx 1$ $\mu$s. On the other hand, in Fig. 2(b) this maximum 
reaches $P_{\upa} \approx 0.25$ $(0.20)$ for $n=1$ $(2)$ and $| \beta | \approx 1.96$ $(3.03)$ with $\kappa \approx 0.24$ 
$(0.23)$ and ${\rm T}_{{\rm t}} \approx 604.5$ $\mu$s (605.5 $\mu$s). Furthermore, note that for $n=2$ and $| \beta | 
\approx 1.27$ the probability $P_{\upa}(T)$ is approximately equal to $0.34$ $(0.18)$ when the superposition has $N=4$ 
$(8)$ states. In fact, this value depends on the number of cycles involved in the sequence and decreases when 
superpositions with $N \gg 1$ are regarded (compare these values with that obtained in Fig. 1). The explanation of this 
result is associated with the quantum interference effect between the $2^{\ell + 1}$ components of the motional state 
$| \Psi_{n}^{(2^{\ell + 1})}(\beta) \rg$, i.e., high values of $N$ lead us to obtain a large number of components 
interfering with each other, and this interference decreases the value of $P_{\upa}(T)$.
\bfig[!t]
\centering
\begin{minipage}[b]{0.45\linewidth}
\includegraphics[width=\linewidth]{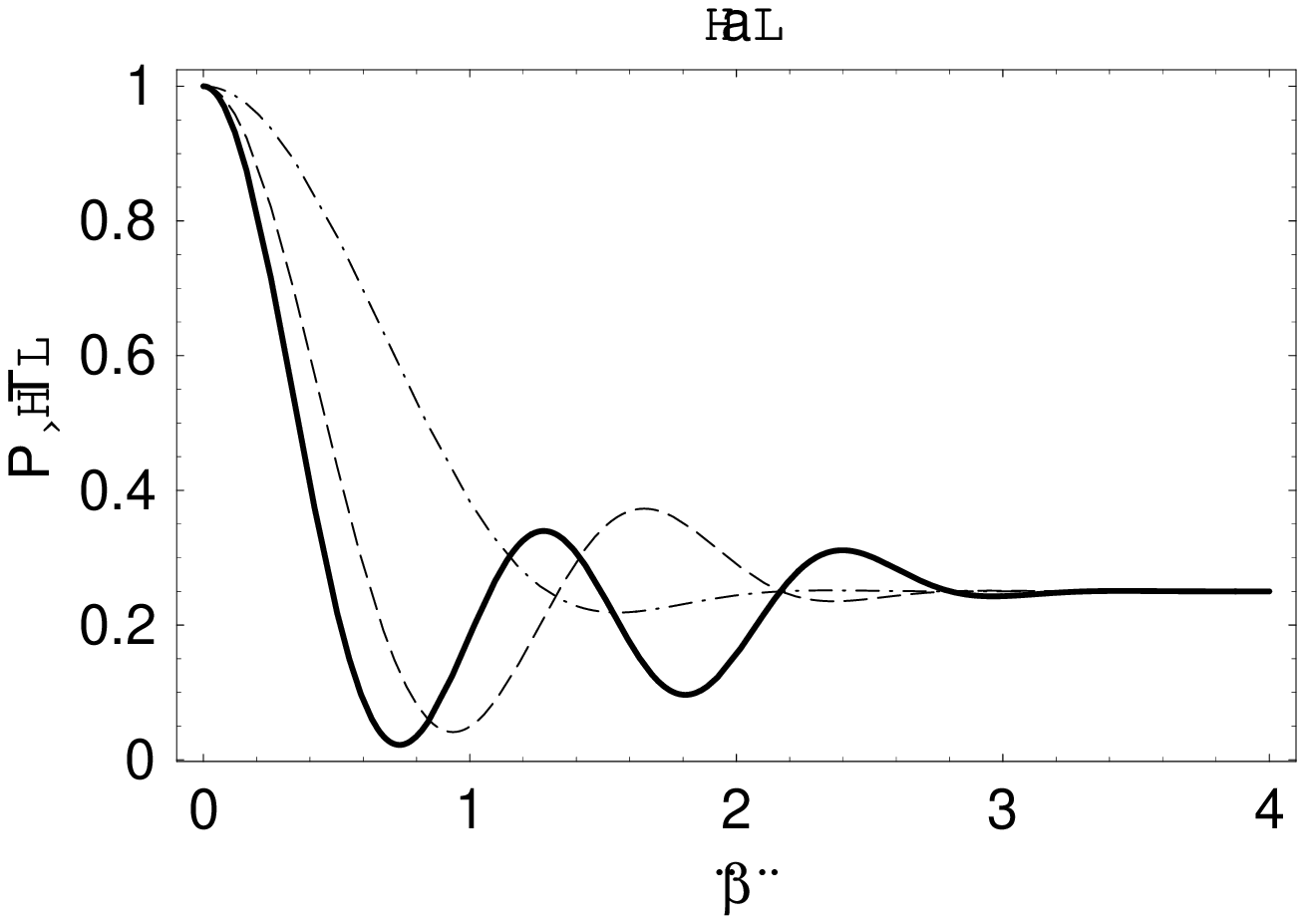}
\end{minipage} \hfill
\begin{minipage}[b]{0.45\linewidth}
\includegraphics[width=\linewidth]{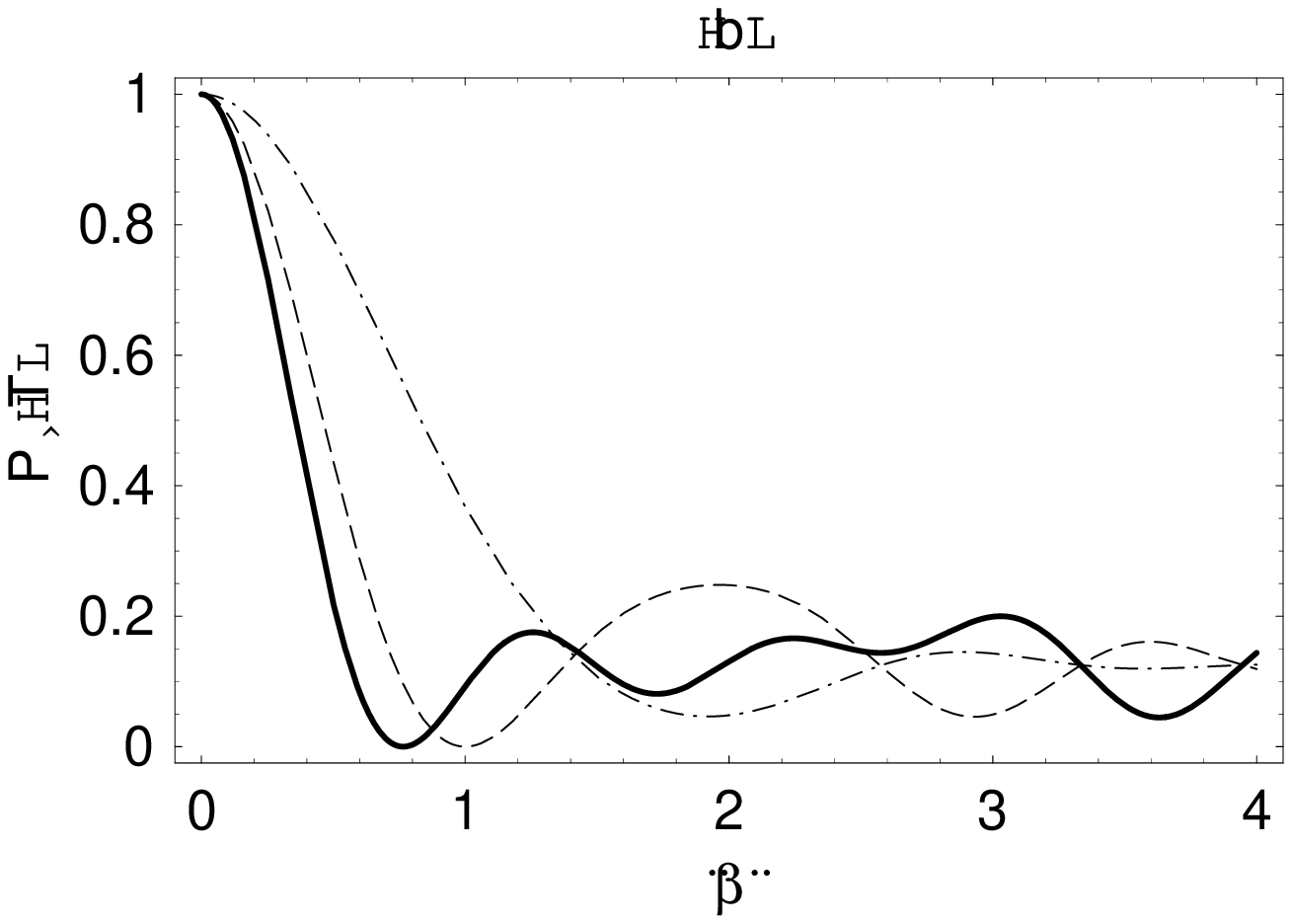}
\end{minipage}
\caption{Plot of $P_{\upa}(T)$ versus $| \beta | \in [0,4]$ for superpositions involving (a) $N=4$ (one cycle) and (b) 
$N=8$ (two cycles) displaced number states. The total duration of laser pulses employed in both sequences for each 
excitation degree $n$ is much more than that proposed by Ref. \cite{r44}, where the number states were generated through a
sequence of quantum nondemolition measurements on a thermal and a coherent initial state. Here, the dot-dashed, dashed and 
solid lines correspond to $n=0,1$ and $2$, respectively.}
\efig

\section{The degradation of the quantum interference effects via Wigner function}

In this section, we adopt the procedure established by Refs. \cite{r16,r42} and assume that the effective relaxation 
process of trapped ions can be described in the framework of the standard master equation for the damped harmonic 
oscillator \cite{r45},
\be
\lb{e14}
\frac{d \ro_{v}}{dt} = - \im \omega_{0} \lbk {\bf a}^{\dagger} {\bf a},\ro_{v} \rbk + \gamma (\bar{n}+1) \lpar 2 {\bf a}
\ro_{v} {\bf a}^{\dagger} - {\bf a}^{\dagger} {\bf a} \ro_{v} - \ro_{v} {\bf a}^{\dagger} {\bf a} \rpar + \gamma \bar{n}
\lpar 2 {\bf a}^{\dagger} \ro_{v} {\bf a} - {\bf a} {\bf a}^{\dagger} \ro_{v} - \ro_{v} {\bf a} {\bf a}^{\dagger} \rpar 
\; ,
\ee
where ${\bf a}$ $({\bf a}^{\dagger})$ is the annihilation (creation) operator associated with the oscillatory motion of
frequency $\omega_{0}$ in a one-dimensional harmonic trap, $\ro_{v}$ describes the density operator for the CM motional 
state at time $t$, $\bar{n}$ is the equilibrium mean number of motional quanta in the reservoir, and $\gamma$ is a positive 
relaxation rate of the energy to thermal equilibrium (the connection between master equation and averaged interferometer 
approach in the trapped ion context was addressed in Ref. \cite{r25}). If one considers the Wigner representation, this 
equation is equivalent to the Fokker-Planck equation,
\be
\lb{e15}
\frac{\partial W}{\partial t} = \lbk \frac{\partial}{\partial q} (\gamma q - \omega_{0} p) + \frac{\partial}{\partial p} 
(\gamma p + \omega_{0} q) + \gamma (\bar{n} + 1/2) \lpar \frac{\partial^{2}}{\partial q^{2}} + \frac{\partial^{2}}
{\partial p^{2}} \rpar \rbk W \; ,
\ee
for the time-dependent Wigner function $W(p,q;t)$, whose solution can be written as an integral equation \cite{r36}
\be
\lb{e16}
W(p,q;t) = \int_{-\infty}^{\infty} \mathbb{K}(p,q;t|p',q';0) W(p',q';0) \; {\rm d} \Gamma' \qquad 
\lpar {\rm d} \Gamma' \equiv dp' dq' \rpar
\ee
with the kernel
\bd
\mathbb{K}(p,q;t|p',q';0) = \lbk \pi (1 + 2 \bar{n})u \rbk^{-1} \exp \lbr - \lbk (1 + 2 \bar{n})u \rbk^{-1} \lbk \lpar 
{\rm p}_{t} - e^{- \gamma t} p' \rpar^{2} + \lpar {\rm q}_{t} - e^{- \gamma t} q' \rpar^{2} \rbk \rbr
\ed
depending on the time variable and reservoir parameters, ${\rm p}_{t} = p \cos (\omega_{0} t) + q \sin (\omega_{0} t)$,
${\rm q}_{t} = q \cos (\omega_{0} t) - p \sin (\omega_{0} t)$, and $u(t) = 1 - e^{- 2 \gamma t}$ (this function was
denominated as `compact time' in Refs. \cite{r16,r36}).

In order to calculate the time evolution of the Wigner function associated to the motional state $| \Psi_{n}^{(2^{\ell + 
1})} (\beta) \rg$, firstly we substitute into the integrand of Eq. (\ref{e16}) the initial Wigner function \cite{r34}
\br
\lb{e17}
W_{n}(p',q';0) &=& \frac{(-1)^{n}}{\pi} \; | \norm_{n}^{(2^{\ell + 1})} |^{2} \lbr \sum_{r=1}^{2^{\ell + 1}} \exp \lpar - 
{\frak R}_{rr} \rpar L_{n} \lpar 2 {\frak R}_{rr} \rpar \right. \nn \\
& & + \; 2 \left. \sum_{s=1}^{2^{\ell + 1}-1} \sum_{r=s+1}^{2^{\ell + 1}} \exp \lbk - \re \lpar {\frak R}_{rs} \rpar \rbk
\cos \lbk \ima \lpar {\frak R}_{rs} \rpar \rbk L_{n} \lbk 2 \re \lpar {\frak R}_{rs} \rpar \rbk \rbr
\er
with
\bd
{\frak R}_{rs}(0) = \lbk (q' + \im p') - \sqrt{2} \; \beta_{r} \rbk \lbk (q' + \im p') - \sqrt{2} \; \beta_{s} \rbk^{\ast} 
+ | \beta |^{2} - \beta_{r} \beta_{s}^{\ast} \; .
\ed
Then, carrying out the integrations in the variables $p'$ and $q'$, we get
\br
\lb{e18}
W_{n}(p,q;t) &=& \frac{(-1)^{n}}{\pi} \frac{| \norm_{n}^{(2^{\ell + 1})} |^{2}}{1 + 2 \bar{n} u} \lbk \frac{1-2(\bar{n}+1)u}
{1 + 2 \bar{n} u} \rbk^{n} \lbr \sum_{r=1}^{2^{\ell + 1}} \exp \lpar - {\frak F}_{rr} \rpar L_{n} \lpar 2 {\frak G}_{rr}
\rpar \right. \nn \\
& & + \; 2 \left. \sum_{s=1}^{2^{\ell + 1}-1} \sum_{r=s+1}^{2^{\ell + 1}} \re \lbk \exp \lpar - {\frak F}_{rs} \rpar L_{n}
\lpar 2 {\frak G}_{rs} \rpar \rbk \rbr \; ,
\er
where
\br
\re \lbk {\frak F}_{rs}(t) \rbk &=& \frac{\re \lbk {\frak R}_{rs}(t) \rbk}{1 + 2 \bar{n} u} + \frac{2 [(1 + 2 \bar{n}) u] 
| \beta |^{2}}{1 + 2 \bar{n} u} \sin^{2} \lbk \frac{\pi (r-s)}{2^{\ell + 1}} \rbk \; , \nn \\
\ima \lbk {\frak F}_{rs}(t) \rbk &=& - \frac{\ima \lbk {\frak R}_{rs}(t) \rbk}{1 + 2 \bar{n} u} + \frac{[(1 + 2 \bar{n}) u] 
| \beta |^{2}}{1 + 2 \bar{n} u} \sin \lbk \frac{2 \pi (r-s)}{2^{\ell + 1}} \rbk \; , \nn \\
\re \lbk {\frak G}_{rs}(t) \rbk &=& \frac{(1-u) \re \lbk {\frak R}_{rs}(t) \rbk}{(1 + 2 \bar{n} u)[1 - 2(\bar{n}+1)u]} -
\frac{2 [(1 + 2 \bar{n}) u]^{2} | \beta |^{2}}{(1 + 2 \bar{n} u)[1 - 2(\bar{n}+1)u]} \sin^{2} \lbk \frac{\pi (r-s)}
{2^{\ell + 1}} \rbk \; , \nn \\
\ima \lbk {\frak G}_{rs}(t) \rbk &=& \frac{[(1 + 2 \bar{n}) u] \ima \lbk {\frak R}_{rs}(t) \rbk}{(1 + 2 \bar{n} u)
[1 - 2(\bar{n}+1)u]} + \frac{(1-u)[(1 + 2 \bar{n}) u] | \beta |^{2}}{(1 + 2 \bar{n} u)[1 - 2(\bar{n}+1)u]} \sin \lbk
\frac{2 \pi (r-s)}{2^{\ell + 1}} \rbk \; , \nn   
\er
and
\bd
{\frak R}_{rs}(t) = \lbk ({\rm q}_{t} + \im {\rm p}_{t}) - \sqrt{2(1-u)} \; \beta_{r} \rbk \lbk ({\rm q}_{t} + \im 
{\rm p}_{t}) - \sqrt{2(1-u)} \; \beta_{s} \rbk^{\ast} + \lpar | \beta |^{2} - \beta_{r} \beta_{s}^{\ast} \rpar (1-u) \; .
\ed
Note that $W_{n}(p,q;t)$ is factorized into diagonal and nondiagonal terms. This permits us, in particular, to investigate 
the degradation of the quantum interference effects among the $2^{\ell + 1}$ components of the motional state represented
by the initial Wigner function $W_{n}(p,q;0)$. Similarly, Chountasis and Vourdas \cite{r46,r47} have employed the same
factorization for the Weyl and Wigner functions associated with a superposition of $m$ quantum states $| s_{i} \rg$, and 
showed that the nondiagonal terms describe the interference effects between the states $| s_{i} \rg$. To illustrate these
results, in Fig. 3(a) we have plotted the three-dimensional picture of $W_{n}(p,q;0)$ versus $p$ and $q$ (containing both
diagonal and nondiagonal terms) for $n=2$, $| \beta | = 3.03$, and $\ell = 2$ fixed; while (c) and (e) correspond to 
diagonal and nondiagonal terms, respectively. The influence of these terms on the shape of the Wigner function (\ref{e17}) 
leads us to confirm the results previously obtained by Chountasis and Vourdas since the nondiagonal term (e) is responsible 
for the interference pattern observed in (a). Now, the degradation of this pattern for $t > 0$ is connected with the 
effective relaxation process under consideration. To illustrate this point, we have plotted in Fig. 3(b) the 
three-dimensional picture of $W_{n}(p,q;t)$ for the same parameter set used in the previous figure, with addition of 
$\omega_{0} / \gamma = 1$, $\bar{n}=1$ and $\gamma t = 0.1$. Furthermore, Figs. 3(d) and (f) represent the diagonal and 
nondiagonal terms, respectively. From the comparison between Figs. 3(a) and (b) we can perceive that the quantum 
inteference pattern present in the first picture has disappeared in (b), and this fact is associated with the decoherence 
effect on the nondiagonal elements of the density operator $\ro_{v}(t)$ (here mapped into the nondiagonal term of 
$W_{n}(p,q;t)$ and pictured through figure 3(f)). It is important mentioning that (d) also has a significant contribution 
to the shape of (b), and when $\gamma t \approx 1$ this contribution is dominant if one compares with that obtained from 
the nondiagonal term. Similar results can be reached if $\gamma t = 0.1$ and $\bar{n} \gg 1$, since the equilibrium mean 
number of quanta in the reservoir represents a scale factor for the compact time $u(t)$. 
\bfig[!t]
\centering
\begin{minipage}[b]{0.45\linewidth}
\includegraphics[width=\linewidth]{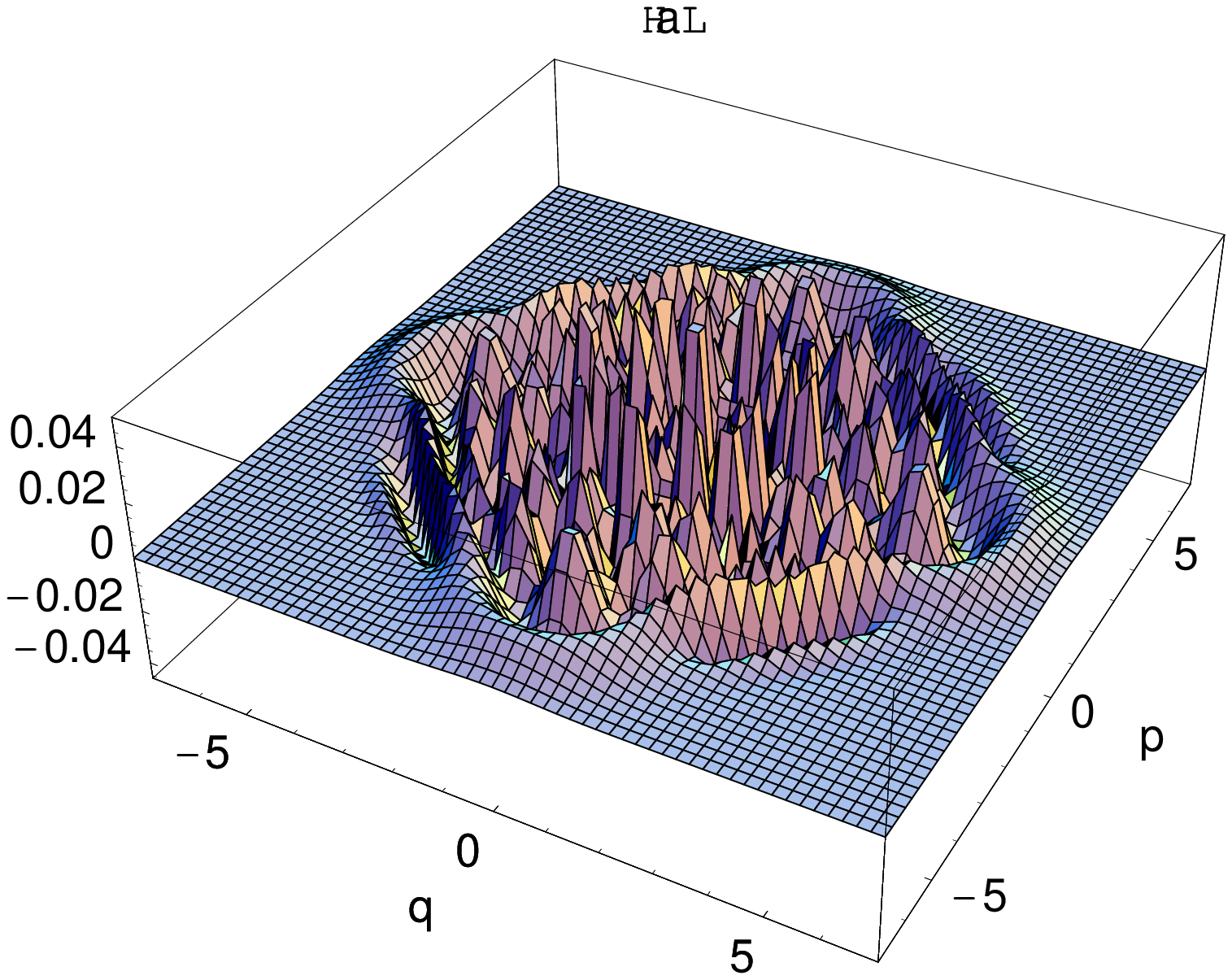}
\end{minipage} \hfill
\begin{minipage}[b]{0.45\linewidth}
\includegraphics[width=\linewidth]{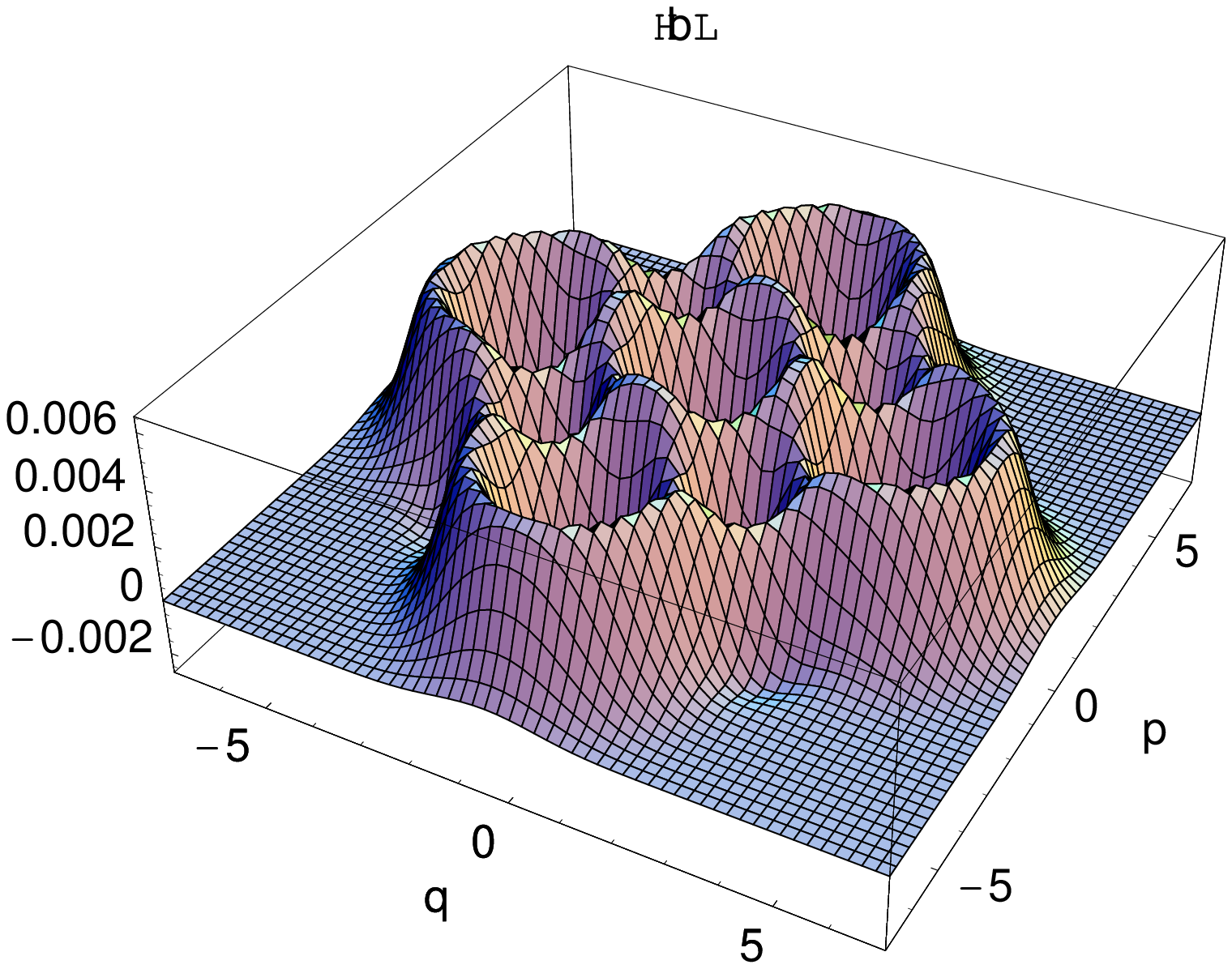}
\end{minipage} \hfill
\begin{minipage}[b]{0.45\linewidth}
\includegraphics[width=\linewidth]{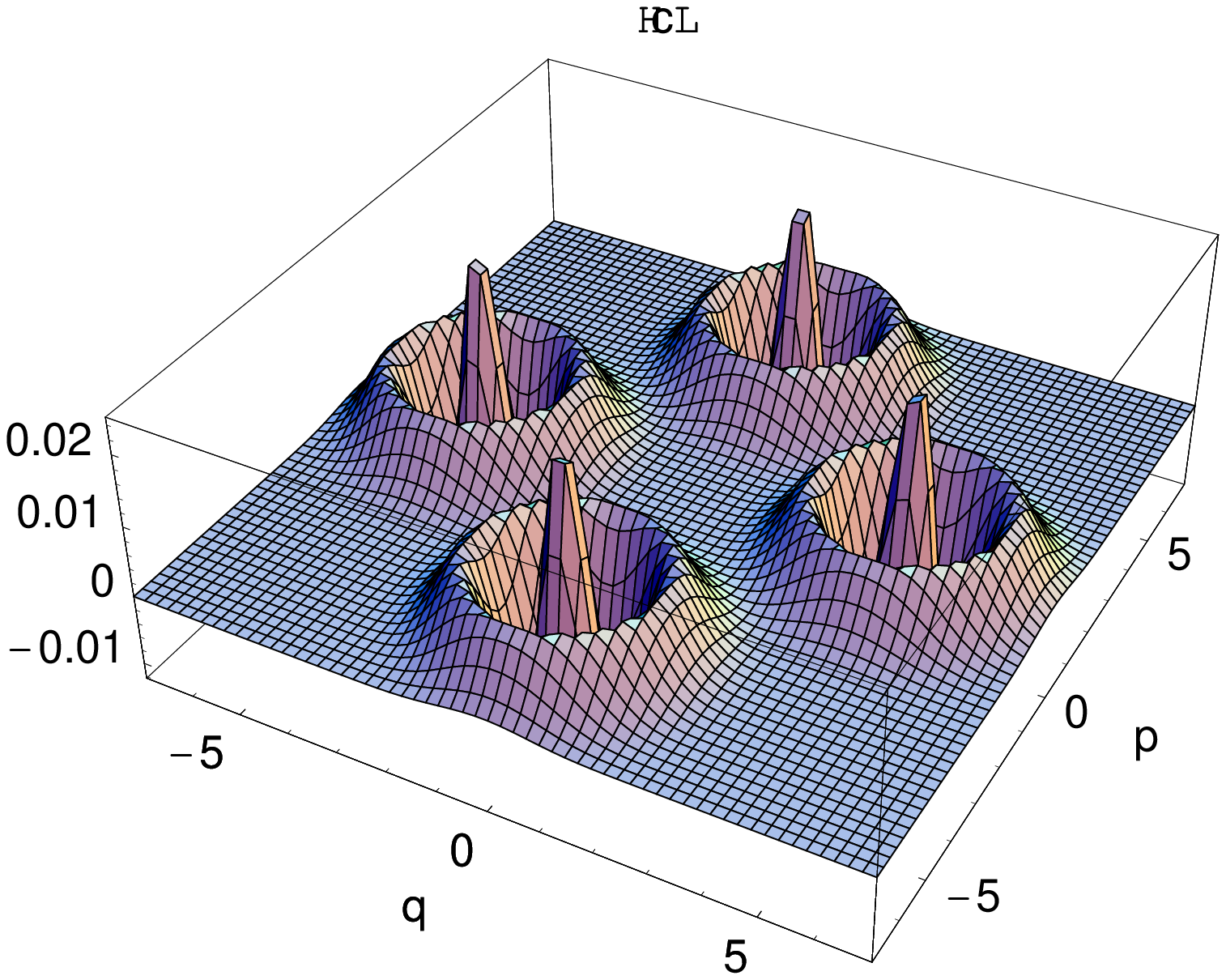}
\end{minipage} \hfill
\begin{minipage}[b]{0.45\linewidth}
\includegraphics[width=\linewidth]{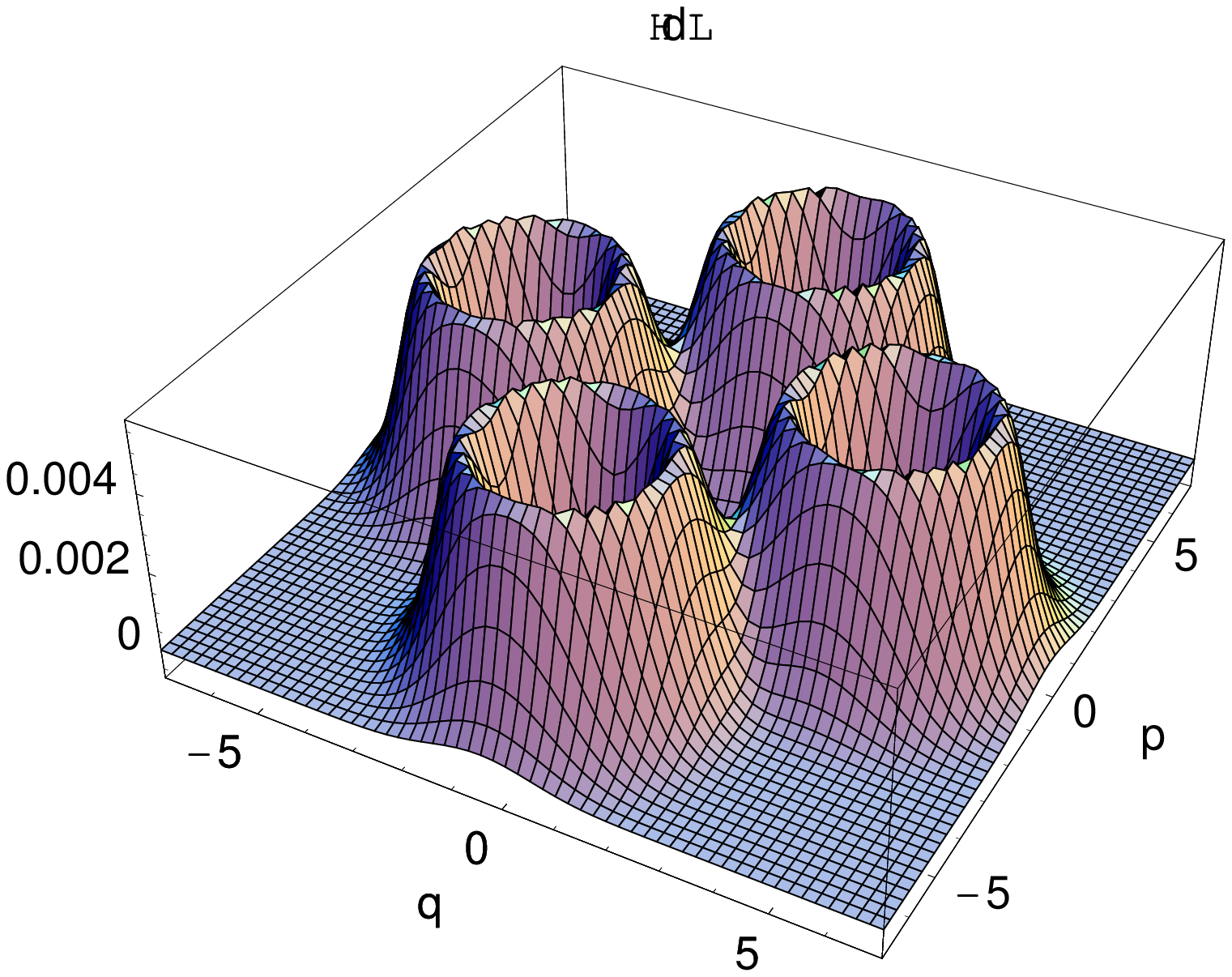}
\end{minipage} \hfill
\begin{minipage}[b]{0.45\linewidth}
\includegraphics[width=\linewidth]{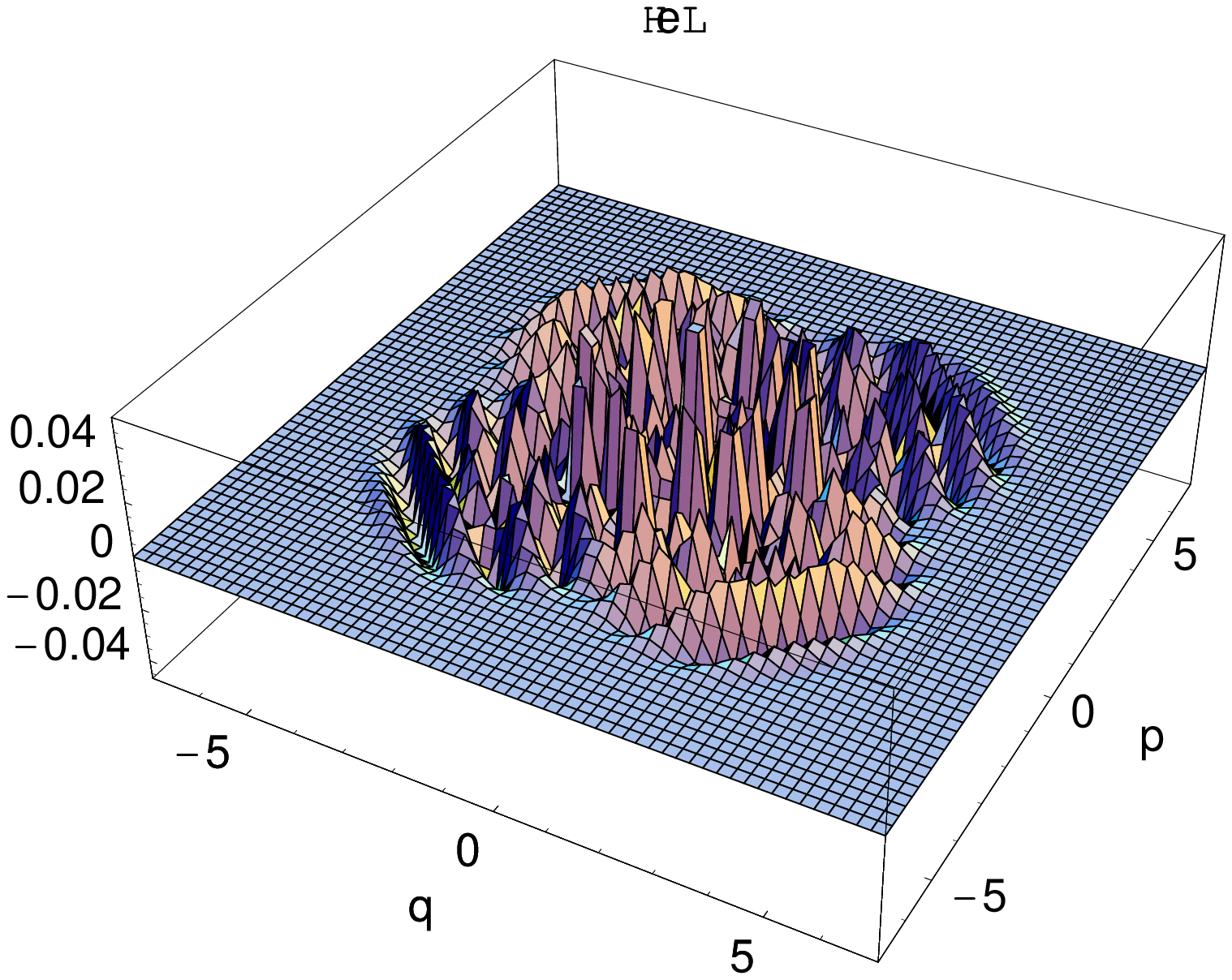}
\end{minipage} \hfill
\begin{minipage}[b]{0.45\linewidth}
\includegraphics[width=\linewidth]{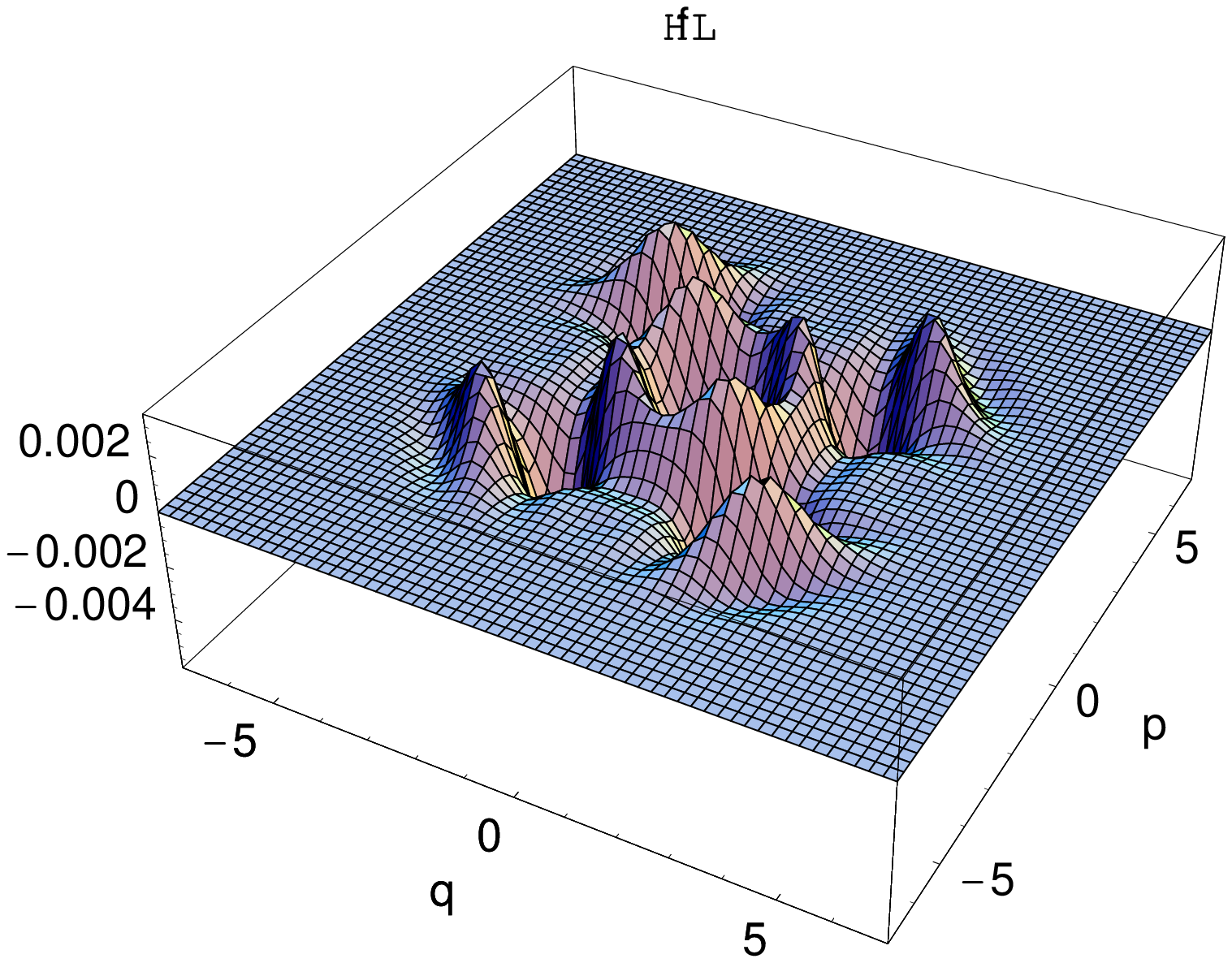}
\end{minipage}
\caption{The pictures (a,c,e) represent the three-dimensional plot of $W_{n}(p,q;0)$ versus $p$ and $q$ for $n=2$, 
$| \beta | = 3.03$, and $\ell = 2$; while (b,d,f) correspond to $W_{n}(p,q;t)$ with $\omega_{0} / \gamma = 1$, $\bar{n}=1$,
and $\gamma t = 0.1$. The interference pattern observed in (a) (containing both diagonal and nondiagonal terms) is a direct 
consequence of the nondiagonal term (e), and its disappearance in (b) is connected with the effective relaxation process
under consideration.}
\efig

\bfig[!t]
\centering
\begin{minipage}[b]{0.45\linewidth}
\includegraphics[width=\linewidth]{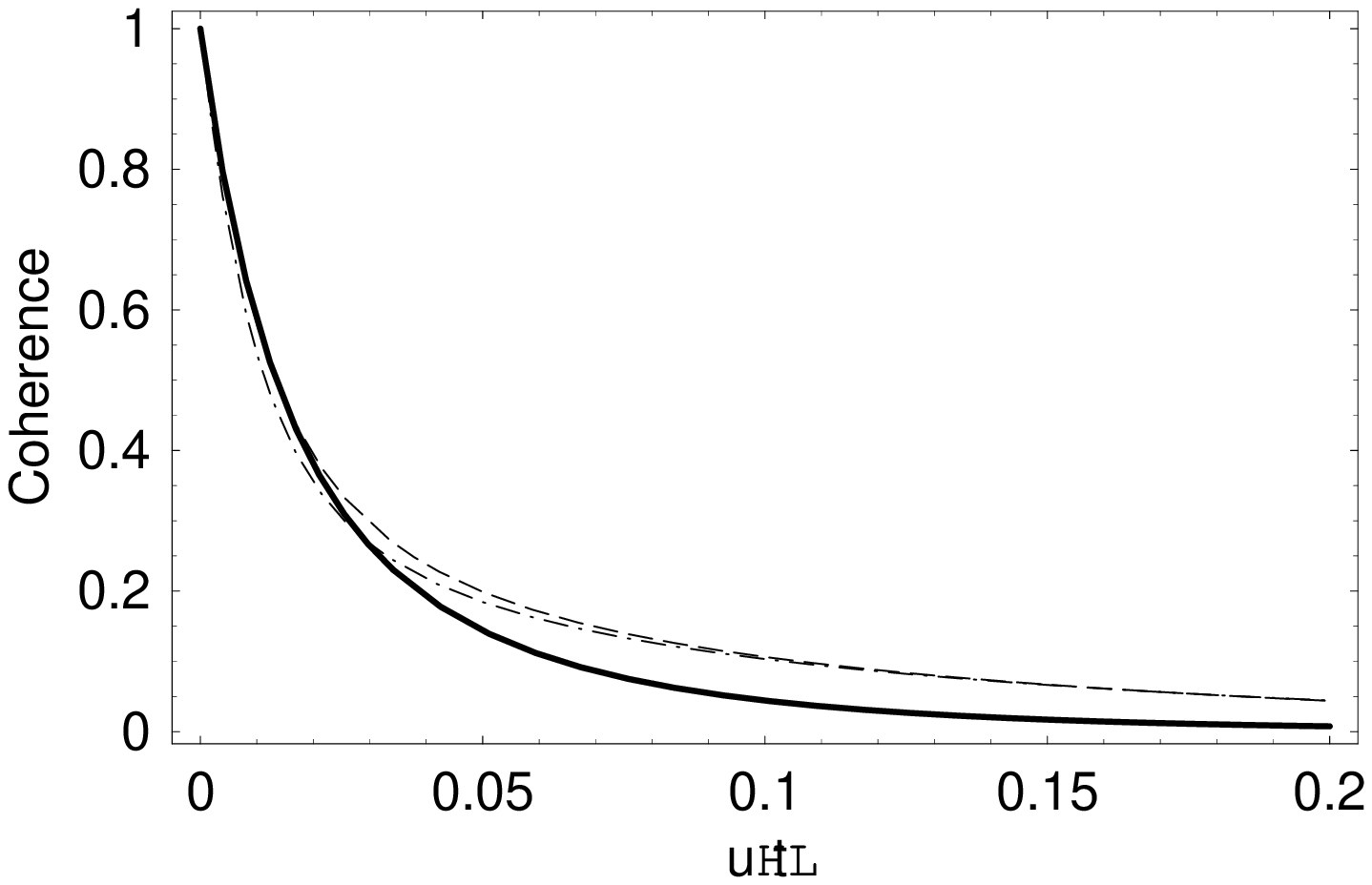}
\end{minipage}
\caption{Plot of ${\cal C}_{n}^{(\ell)}(t)$ versus $u(t)$ for $| \beta | = 3.03$, $\ell = 2$, and $\bar{n}=1$ fixed, where
the dot-dashed, dashed, and solid lines correspond to $n=0$, $1$, and $2$, respectively. Note that in a short period of 
time $(\gamma t \in [0,0.1116])$ and high values of $\ell$ and $n$, the measure of coherence suddenly goes to zero.}
\efig
The measure of coherence ${\cal C}(t)$ was introduced in \cite{r36} as a quantitative measurement which characterizes the
rate of decoherence in the Fock-state basis. Since {\em decoherence} can be interpreted as the disappearence, with time
progression, of the nondiagonal elements associated to the density operator $\ro_{v}(t)$, we will use this measure in order 
to explain the pattern observed in the diagonal and nondiagonal terms of $W_{n}(p,q;t)$. For this purpose, we define the 
normalized measure of coherence through the expression (for more details, see Appendix A)
\be
\lb{e19}
{\cal C}_{n}^{(\ell)}(t) = \frac{\mu_{n}^{(\ell)}(t) - \lam_{n}^{(\ell)}(t)}{\mu_{n}^{(\ell)}(0) - \lam_{n}^{(\ell)}(0)} 
\; ,
\ee
where the `total purity' $\mu_{n}^{(\ell)}(t)$ and `diagonal purity' $\lam_{n}^{(\ell)}(t)$ can be written in terms of the
initial Wigner function as follows:
\br
\lb{e20}
\mu_{n}^{(\ell)}(t) &=& \int_{- \infty}^{\infty} \int_{- \infty}^{\infty} \Xi (p',q',p'',q'';t) W_{n}(p',q';0) 
W_{n}(p'',q'';0) \; {\rm d} \Gamma' {\rm d} \Gamma'' \; , \\
\lb{e21}
\lam_{n}^{(\ell)}(t) &=& \int_{- \infty}^{\infty} \int_{- \infty}^{\infty} \Delta (p',q',p'',q'';t) W_{n}(p',q';0) 
W_{n}(p'',q'';0) \; {\rm d} \Gamma' {\rm d} \Gamma'' \; ,
\er
being $\Xi$ and $\Delta$ given by Eqs. (\ref{a7}) and (\ref{a10}), respectively. The functions $\mu_{n}^{(\ell)}(t)$ and 
$\lam_{n}^{(\ell)}(t)$ were explicitly calculated in Appendix A, and their analytical results for $n=0$ corroborate that
obtained by Souza Silva et al. \cite{r16}. Figure 4 represents the plot of ${\cal C}_{n}^{(\ell)}(t)$ versus the compact
time $u(t) = 1 - e^{- 2 \gamma t}$ for $| \beta | = 3.03$, $\ell = 2$, and $\bar{n}=1$ fixed, where the dot-dashed, dashed,
and solid lines correspond to $n=0$, $1$, and $2$, respectively. Note that for $u=0.2$ we obtain ${\cal C}_{0}^{(2)} \approx
0.0443$, ${\cal C}_{1}^{(2)} \approx 0.0438$, and ${\cal C}_{2}^{(2)} \approx 0.0076$; while for $n=2$, the measure of
coherence goes to ${\cal C}_{2}^{(0)} \approx 0.0428$, ${\cal C}_{2}^{(1)} \approx 0.0247$, and ${\cal C}_{2}^{(2)} \approx
0.0076$ at the same compact time. Otherwise, if one considers $u=1$ in both situations, ${\cal C}_{n}^{(\ell)}$ goes to
zero for any values of $\ell$ and $n$. Thus, high values of the parameters $\ell$ and $n$ considerably decrease (increase)
the measure of coherence (decoherence process) in a short period of time and this fact could explain the disappearance of
the quantum interference pattern observed in Fig. 3 \cite{r48}.

\section{Summary and conclusions}

In this paper, we have combined different theoretical approaches in order to engineer superpositions of displaced number
states on a circle in phase space as target states for the center-of-mass motion of a trapped ion. The total duration $T$
of laser pulses employed in the process and the probability $P_{\upa}(T)$ of getting the ion in the upper electronic state
were explicitly calculated and analyzed. In particular, we have verified that (i) the quantum interference effects among 
the $N=2^{\ell + 1}$ components of the motional state described by Eq. (\ref{e1}) decrease the values of $P_{\upa}(T)$ when
superpositions with $N \gg 1$ are regarded, and (ii) the Lamb-Dicke parameter $\kappa$ essentially depends on the number 
$\ell$ of cycles involved in the sequence and the excitation degree $n$ of the motional state. Furthermore, we have also 
investigated the degradation of the quantum interference effects via the Wigner function and showed that (iii) these 
effects basically depend on the nondiagonal term of the Wigner function at time $t=0$, (iv) the quantum interference 
pattern present in $W_{n}(p,q;0)$ disappears for $t > 0$ and this fact is associated with the decoherence process on the 
nondiagonal elements of the density operator $\ro_{v}(t)$, and (v) high values of the parameters $\ell$ and $n$ increase 
the decoherence process in a short period of time. Summarizing, the work reported here is clearly the product of 
considerable effort and constitutes an marginally original contribution to the wider field of quantum state engineering.

Recently, Lvovsky and Babichev \cite{r49} have synthesized the displaced Fock states of the electromagnetic field by
overlapping the pulsed optical single-photon Fock state with coherent states on a high-reflection beam splitter and
showed its nonclassical properties (such as negativity of the Wigner function and photon number oscillations) through a
complete tomographic reconstruction. However, the nonunitary quantum efficiency of the homodyne detector, the dark counts
of the single-photon detector, and the impurity of the optical mode of the conditionally prepared photon represent 
important restrictions on the preparation and measurement of the Fock state. In this sense, experiments involving trapped 
ions are a great laboratory in the construction process of nonclassical states since the decoherence time is the longest if 
one compares with that obtained from experiments for trapped and travelling nonclassical states of the electromagnetic 
field. In conclusion, we believe that the results obtained in this paper can motivate the generation of new nonclassical 
states in future experiments on trapped ions and to contribute significantly to the study of quantum interference effects 
in different physical contexts.
 
\section*{Acknowledgments}

The author MAM acknowledges the hospitality of the {\sl Departamento de Ci\^{e}ncias Exatas e Tecnol\'{o}gicas} of 
{\sl Universidade Estadual de Santa Cruz} (Ilh\'{e}us, Bahia, Brazil) where this work was initiated. WDJ acknowledges
financial support from PRODOC/FAPEX, Bahia, Brazil, project no. 991042-69. MAM and WDJ are grateful to R.J. Napolitano 
for reading the manuscript and for providing valuable suggestions. This work was supported by FAPESP, S\~{a}o Paulo, Brazil, 
project nos. 01/11209-0 and 00/15084-5.

\appendix
\section{The measure of coherence}

Dodonov et al. \cite{r36} have introduced two quantitative measures which characterize the rates of decoherence and 
thermalization of quantum systems, and studied the time evolution of these measures in the case of a quantum harmonic
oscillator whose relaxation process is described in the framework of the standard master equation. In particular, the 
measure of coherence ${\cal C}(t)$ was defined by the authors through the expression
\be
\lb{a1}
{\cal C}(t) = \frac{\mu(t) - \lam(t)}{\mu(0) - \lam(0)} \; ,
\ee
where the functions $\mu(t) \equiv \tr \lbk \ro^{2}(t) \rbk$ and $\lam(t) \equiv \tr \lbk \ro_{d}^{2}(t) \rbk$ ($\ro_{d}(t)
= \sum_{n \in \mathbb{N}} P_{n}(t) | n \rg \lg n |$ correspond to the diagonal part of the density operator $\ro(t)$, being
$P_{n}(t) = \lg n | \ro(t) | n \rg$) connected with Wigner function $W(p,q;t)$ and phonon (photon) distribution function 
$P_{n}(t)$, as follow:
\be
\lb{a2}
\mu(t) = \int_{- \infty}^{\infty} \lbk W(p,q;t) \rbk^{2} {\rm d} \Gamma \qquad \lpar {\rm d} \Gamma = dp \, dq \rpar
\ee
and
\be
\lb{a3}
\lam(t) = \sum_{n=0}^{\infty} \lbk P_{n}(t) \rbk^{2} \; .
\ee
In addition, the phonon (photon) distribution function can also be obtained by means of the auxiliary relation
\be
\lb{a4}
P_{n}(t) = \int_{- \infty}^{\infty} W_{n}(p,q) W(p,q;t) \; {\rm d} \Gamma \; ,
\ee
being
\bd
W_{n}(p,q) = 2 (-1)^{n} \exp \lbk - \lpar p^{2} + q^{2} \rpar \rbk L_{n} \lbk 2 \lpar p^{2} + q^{2} \rpar \rbk
\ed
the Wigner function associated to the number state at time $t=0$. Thus, if one knows the Wigner function associated to the
density operator $\ro(t)$, then the functions `total purity' $\mu(t)$ and `diagonal purity' $\lam(t)$ can be promptly 
calculated. Note that ${\cal C}(0) = 1$, and ${\cal C}(t) = 0$ for any completely incoherent state without nondiagonal 
matrix elements in the energy basis (provided that initially at least one nondiagonal element was different from zero).

Now, substituting the solution (\ref{e16}) of the Fokker-Planck equation for the harmonic oscillator into Eqs. (\ref{a2}) 
and (\ref{a4}), we obtain
\br
\lb{a5}
\mu(t) &=& \int_{- \infty}^{\infty} \int_{- \infty}^{\infty} \Xi (p',q',p'',q'';t) W(p',q';0) W(p'',q'';0) \; {\rm d} 
\Gamma' {\rm d} \Gamma'' \; , \\
\lb{a6}
P_{n}(t) &=& \int_{- \infty}^{\infty} {\cal K}_{n}(p',q';t) W(p',q';0) \; {\rm d} \Gamma' \; ,
\er
with
\be
\lb{a7}
\Xi (p',q',p'',q'';t) = \lbk (1 + 2 \bar{n})u \rbk^{-1} \exp \lbr - \frac{(1-u) \lbk (p'-p'')^{2} + (q'-q'')^{2} \rbk}
{2(1 + 2 \bar{n})u} \rbr
\ee
and
\br
\lb{a8}
{\cal K}_{n}(p',q';t) &=& \frac{2 (-1)^{n}}{1+(1+2\bar{n})u} \lbk \frac{1-(1+2\bar{n})u}{1+(1+2\bar{n})u} \rbk^{n} \exp 
\lbk - \frac{(1-u) \lpar p'^{2}+q'^{2} \rpar}{1+(1+2\bar{n})u} \rbk \nn \\
& & \times \; L_{n} \lbk \frac{2(1-u) \lpar p'^{2}+q'^{2} \rpar}{[1-(1+2\bar{n})u][1+(1+2\bar{n})u]} \rbk \; .
\er
Consequently, the function $\lam(t)$ can also be determined through the equation
\be
\lb{a9}
\lam(t) = \int_{- \infty}^{\infty} \int_{- \infty}^{\infty} \Delta (p',q',p'',q'';t) W(p',q';0) W(p'',q'';0) \; {\rm d} 
\Gamma' {\rm d} \Gamma'' \; ,
\ee
where $\Delta (p',q',p'',q'';t)$ is given by
\br
\lb{a10}
\Delta(p',q',p'',q'';t) &=& \lbk (1 + 2 \bar{n})u \rbk^{-1} \exp \lbk - \frac{(1-u)(p'^{2} + q'^{2} + p''^{2} + q''^{2})}
{2(1 + 2 \bar{n})u} \rbk \nn \\
& & \times \; I_{0} \lbk \frac{(1-u) \lbk (p'^{2} + q'^{2})(p''^{2} + q''^{2}) \rbk^{1/2}} {(1 + 2 \bar{n})u} \rbk \; ,
\er
being $I_{\nu}(z)$ the modified Bessel function of the first kind \cite{r50}. This expression was obtained with the help of
the intermediate relation \cite{r51}
\bd
\sum_{n=0}^{\infty} \frac{n! L_{n}^{(\nu)}(x) L_{n}^{(\nu)}(y)}{\Gamma(n + \nu + 1)} \, z^{n} = (1-z)^{-1} (xyz)^{-\nu/2} 
\exp \lbk - \frac{z(x+y)}{1-z} \rbk I_{\nu} \lbk \frac{2 (xyz)^{1/2}}{1-z} \rbk \; ,
\ed
with $|z| < 1$ and $\nu > -1$. Since the initial Wigner function does not depend on the undermentioned relaxation process,
Eqs. (\ref{a5}) and (\ref{a9}) represent an alternative way to the calculation of the total and diagonal purities.

In many situations of practical interest both purities can be established rather easily. For instance, if one considers the
initial Wigner function (\ref{e17}) in Eqs. (\ref{a5}) and (\ref{a9}), we get
\be
\lb{a11}
\mu_{n}^{(\ell)}(t) = \half \lbk \frac{2 | \norm_{n}^{(2^{\ell + 1})} |^{2}}{n!} \rbk^{2} \frac{\partial^{2n}}{\partial 
x^{n} \partial y^{n}} \left. \frac{(xy)^{n}}{x {\rm A}_{y} + (1-u)y} \; {\rm H}(x,y;t) \right|_{x=y=1}
\ee
and
\be
\lb{a12}
\lam_{n}^{(\ell)}(t) = \lbk \frac{2 | \norm_{n}^{(2^{\ell + 1})} |^{2}}{n!} \rbk^{2} \frac{\partial^{2n}}{\partial x^{n} 
\partial y^{n}} \left. \frac{(xy)^{n}}{{\cal A}_{+} {\cal B}_{+} - {\cal A}_{-} {\cal B}_{-}} \; {\rm J}(x,y;t) 
\right|_{x=y=1} \; ,
\ee
where
\br
{\rm H}(x,y;t) &=& \sum_{p,q=1}^{2^{\ell + 1}} \sum_{r,s=1}^{2^{\ell + 1}} \exp \lbk - 2 \lpar {\frak D}_{pq}^{rs} - 
{\frak C}_{pq}^{rs} \rpar | \beta |^{2} \rbk \; , \nn \\
{\rm J}(x,y;t) &=& \sum_{p,q=1}^{2^{\ell + 1}} \sum_{r,s=1}^{2^{\ell + 1}} \exp \lbr - 2 \lbk \frac{{\cal B}_{+} 
{\rm A}_{rs} {\rm U}_{rs} + {\cal A}_{+} {\rm B}_{pq} {\rm V}_{pq}}{{\cal A}_{+} {\cal B}_{+}} + \frac{2(1-u) \lpar 
{\cal B}_{+} {\cal B}_{-} {\rm U}_{rs}^{2} + {\cal A}_{+} {\cal A}_{-} {\rm V}_{pq}^{2} \rpar}{{\cal A}_{+} {\cal B}_{+} 
\lpar {\cal A}_{+} {\cal B}_{+} - {\cal A}_{-} {\cal B}_{-} \rpar} \rbk | \beta |^{2} \rbr \nn \\
& & \times \; I_{0} \lbk \frac{8(1-u) {\rm U}_{rs} {\rm V}_{pq} | \beta |^{2}}{{\cal A}_{+} {\cal B}_{+} - {\cal A}_{-}
{\cal B}_{-}} \rbk \; , \nn
\er
with
\br
{\frak C}_{pq}^{rs}(x,y;t) &=& \lbk x {\rm A}_{y} + (1-u)y \rbk^{-1} \lbr {\rm A}_{y} {\rm U}_{rs}^{2} + 2(1-u) \cos \lbk 
\frac{\pi (r+s-p-q)}{2^{\ell + 1}} \rbk {\rm U}_{rs} {\rm V}_{pq} + {\rm A}_{x} {\rm V}_{pq}^{2} \rbr \; , \nn \\
{\frak D}_{pq}^{rs}(x,y;t) &=& \cos \lbk \frac{\pi (r-s)}{2^{\ell +1}} \rbk {\rm U}_{rs} + \cos \lbk \frac{\pi (p-q)}
{2^{\ell + 1}} \rbk {\rm V}_{pq} \; , \nn \\
{\rm U}_{rs}(x) &=& \cos \lbk \frac{\pi (r-s)}{2^{\ell + 1}} \rbk x + \im \sin \lbk \frac{\pi (r-s)}{2^{\ell + 1}} \rbk 
\; , \nn \\
{\rm V}_{pq}(y) &=& \cos \lbk \frac{\pi (p-q)}{2^{\ell + 1}} \rbk y + \im \sin \lbk \frac{\pi (p-q)}{2^{\ell + 1}} \rbk 
\; , \nn \\
{\rm A}_{rs}(t) &=& \cos \lbk \frac{\pi (r-s)}{2^{\ell + 1}} \rbk (1-u) - \im \sin \lbk \frac{\pi (r-s)}{2^{\ell + 1}} \rbk 
[1+(1+2\bar{n})u] \; , \nn \\
{\rm B}_{pq}(t) &=& \cos \lbk \frac{\pi (p-q)}{2^{\ell + 1}} \rbk (1-u) - \im \sin \lbk \frac{\pi (p-q)}{2^{\ell + 1}} \rbk 
[1+(1+2\bar{n})u] \; , \nn \\
{\cal A}_{\pm}(x;t) &=& [ 1 \pm (1+2\bar{n})u ] x \pm (1-u) \; , \nn \\
{\cal B}_{\pm}(y;t) &=& [ 1 \pm (1+2\bar{n})u ] y \pm (1-u) \; , \nn \\
{\rm A}_{x}(t) &=& 2(1+2\bar{n})u x + (1-u) \; , \nn \\
{\rm A}_{y}(t) &=& 2(1+2\bar{n})u y + (1-u) \; . \nn
\er
Furthermore, the phonon distribution function is given by
\be
\lb{a13}
P_{nm}^{(\ell)}(t) = 2 | \norm_{n}^{(2^{\ell + 1})} |^{2} \, \frac{(-1)^{n+m}}{n!} \frac{\partial^{n}}{\partial x^{n}} 
\left. \frac{{\cal A}_{-}^{m} x^{n}}{{\cal A}_{+}^{m+1}} \, {\rm I}_{m}(x;t) \right|_{x=1} \; ,
\ee
where
\bd
{\rm I}_{m}(x;t) = \sum_{r,s=1}^{2^{\ell + 1}} \exp \lpar - \frac{2 {\rm A}_{rs} {\rm U}_{rs} | \beta |^{2}}{{\cal A}_{+}}
\rpar L_{m} \lbk \frac{4(1-u) {\rm U}_{rs}^{2} | \beta |^{2}}{{\cal A}_{+} {\cal A}_{-}} \rbk \; .
\ed
It is important mentioning that Eqs. (\ref{a11})-(\ref{a13}) were calculated by means of the parametric representation for
the associated Laguerre polynomial \cite{r52}, i.e.,
\bd
L_{n}^{(\alf)}(z) = e^{z} \left. \frac{1}{n!} \frac{d^{n}}{dx^{n}} \, x^{n + \alf} e^{- xz} \right|_{x=1} \; .
\ed
The total and diagonal purities determined in this appendix corroborate that obtained by Souza Silva et al. \cite{r16}
for $n=0$.


\begin{references}
\bib{r1} V.V. Dodonov, `Nonclassical' states in quantum optics: a `squeezed' review of the first 75 years, J. Opt. B: 
Quantum Semiclass. Opt. 4 (2002) R1, and references therein.
%
\bib{r2} C. Monroe, D.M. Meekhof, B.E. King, D.J. Wineland, A `Schr\"{o}dinger cat' superposition state of an atom, Science 
272 (1996) 1131.
%
\bib{r3} D.M. Meekhof, C. Monroe, B.E. King, W.M. Itano, D.J. Wineland, Generation of nonclassical motional states of a 
trapped atom, Phys. Rev. Lett. 76 (1996) 1796.
%
\bib{r4} D. Leibfried, D.M. Meekhof, B.E. King, C. Monroe, W.M. Itano, D.J. Wineland, Experimental determination of the
motional quantum state of a trapped atom, Phys. Rev. Lett. 77 (1996) 4281.
%
\bib{r5} W.M. Itano, C. Monroe, D.M. Meekhof, D. Leibfried, B.E. King, D.J. Wineland, Quantum harmonic oscillator state
syntesis and analysis, SPIE Proc. 2995 (1997) 43.
%
\bib{r6} R.L. de Matos Filho, W. Vogel, Nonlinear coherent states, Phys. Rev. A 54 (1996) 4560.
%
\bib{r7} S.A. Gardiner, J.I. Cirac, P. Zoller, Nonclassical states and measurement of general motional observables of a
trapped ion, Phys. Rev. A 55 (1997) 1683.
%
\bib{r8} C.C. Gerry, Generation of Schr\"{o}dinger cats and entangled coherent states in the motion of a trapped ion by 
a dispersive interaction, Phys. Rev. A 55 (1997) 2478.
%
\bib{r9} S.-B. Zheng, G.-C. Guo, Generation of superpositions of coherent states of the motion of a trapped ion, Eur. Phys. 
J. D 1 (1998) 105.
%
\bib{r10} H. Moya-Cessa, S. Wallentowitz, W. Vogel, Quantum-state engineering of a trapped ion by coherent-state
superpositions, Phys. Rev. A 59 (1999) 2920.
%
\bib{r11} W.D. Jos\'{e}, S.S. Mizrahi, Generation of circular states and Fock states in a trapped ion, J. Opt. B: Quantum 
Semiclass. Opt. 2 (2000) 306.
%
\bib{r12} V. Man'ko, G. Marmo, A. Porzio, S. Solimeno, F. Zaccaria, Trapped ions in laser fields: A benchmark for deformed 
quantum oscillators, Phys. Rev. A 62 (2000) 053407.
%
\bib{r13} G. Huyet, S. Franke-Arnold, S.M. Barnett, Superposition states at finite temperature, Phys. Rev. A 63 (2001)
043812.
%
\bib{r14} Z. Kis, W. Vogel, L. Davidovich, Nonlinear coherent states of trapped-atom motion, Phys. Rev. A 64 (2001) 033401.
%
\bib{r15} M. Feng, Preparation of Schr\"{o}dinger cat states with cold ions beyond the Lamb-Dicke limit, Phys. Lett. A 282 
(2001) 230.
%
\bib{r16} A.L.S. Silva, W.D. Jos\'{e}, V.V. Dodonov, S.S. Mizrahi, Production of two-Fock states superpositions from even 
circular states and their decoherence, Phys. Lett. A 282 (2001) 235.
%
\bib{r17} S.-B. Zheng, Preparation of arbitrary finite superpositions of Fock states for the center-of-mass motion of two 
trapped ions, J. Opt. B: Quantum Semiclass. Opt. 3 (2001) 328.
%
\bib{r18} F.L. Semi\~{a}o, A. Vidiella-Barranco, J.A. Roversi, Nonclassical effects in cold trapped ions inside a cavity, 
Phys. Rev. A 66 (2002) 063403.
%
\bib{r19} N.B. An, T.M. Duc, Generation of three-mode nonclassical vibrational states of ions, Phys. Rev. A 66 (2002)
065401.
%
\bib{r20} S. Schneider, G.J. Milburn, Decoherence in ion traps due to laser intensity and phase fluctuations, Phys. Rev. A 
57 (1998) 3748.
%
\bib{r21} R.M. Serra, N.G. de Almeida, W.B. da Costa, M.H.Y. Moussa, Decoherence in trapped ions due to polarization of 
the residual background gas, Phys. Rev. A 64 (2001) 033419.
%
\bib{r22} A.A. Budini, R.L. de Matos Filho, N. Zagury, Decoherence in non-classical states of a trapped ion, J. Opt. B: 
Quantum Semiclass. Opt. 4 (2002) S462.
%
\bib{r23} A.A. Budini, R.L. de Matos Filho, N. Zagury, Localization and dispersivelike decoherence in vibronic states 
of a trapped ion, Phys. Rev. A 65 (2002) 041402(R).
%
\bib{r24} J.F. Poyatos, J.I. Cirac, P. Zoller, Quantum reservoir engineering with laser cooled trapped ions, Phys. Rev. 
Lett. 77 (1996) 4728.
%
\bib{r25} Q.A. Turchette, C.J. Myatt, B.E. King, C.A. Sackett, D. Kielpinski, W.M. Itano, C. Monroe, D.J. Wineland,
Decoherence and decay of motional quantum states of a trapped atom coupled to engineered reservoirs, Phys. Rev. A 62 (2000)
053807; \\
C.J. Myatt, B.E. King, Q.A. Turchette, C.A. Sackett, D. Kielpinski, W.M. Itano, C. Monroe, D.J. Wineland, Decoherence of
quantum superpositions through coupling to engineered reservoirs, Nature 403 (2000) 269; \\
Q.A. Turchette, D. Kielpinski, B.E. King, D. Leibfried, D.M. Meekhof, C.J. Myatt, M.A. Rowe, C.A. Sackett, C.S. Wood, W.M.
Itano, C. Monroe, D.J. Wineland, Heating of trapped ion from the quantum ground state, Phys. Rev. A 61 (2000) 063418. 
%
\bib{r26} J.I. Cirac, P. Zoller, Quantum computations with cold trapped ions, Phys. Rev. Lett. 74 (1995) 4091.
%
\bib{r27} C. Monroe, D.M. Meekhof, B.E. King, W.M. Itano, D.J. Wineland, Demonstration of a fundamental quantum logic gate, 
Phys. Rev. Lett. 75 (1995) 4714.
%
\bib{r28} D. Jonathan, M.B. Plenio, P.L. Knight, Fast quantum gates for cold trapped ions, Phys. Rev. A 62 (2000) 042307, 
and references therein.
%
\bib{r29} D. Kielpinski, A. Ben-Kish, J. Britton, V. Meyer, M.A. Rowe, C.A. Sackett, W.M. Itano, C. Monroe, D.J. Wineland,
Recent results in trapped-ion quantum computing at NIST, quant-ph/0102086, 2001.
%
\bib{r30} D. Jonathan, M.B. Plenio, Light-shift-induced quantum gates for ions in thermal motion, quant-ph/0103140, 2001.
%
\bib{r31} M. Feng, X. Wang, Implementation of quantum gates and preparation of entangled states in cavity QED with cold 
trapped ions, quant-ph/0112031 v2, 2002.
%
\bib{r32} M. Feng, Quantum computing with trapped ions in an optical cavity via Raman transition, Phys. Rev. A 66 (2002)
054303.
%
\bib{r33} D.J. Wineland, M. Barret, J. Britton, J. Chiaverini, B. DeMarco, W.M. Itano, B. Jelenkovi\'{c}, C. Langer, D.
Leibfried, V. Meyer, T. Rosenband, T. Sch\"{a}tz, Quantum information processing with trapped ions, quant-ph/0212079 v1,
2002, and references therein.
%
\bib{r34} M.A. Marchiolli, L.F. da Silva, P.S. Melo, C.M.A. Dantas, Quantum-interference effects on the superposition of
$N$ displaced number states, Physica A 291 (2001) 449.
%
\bib{r35} S. Wallentowitz, W. Vogel, Reconstruction of the Quantum Mechanical State of a Trapped Ion, Phys. Rev. Lett. 75 
(1995) 2932.
%
\bib{r36} V.V. Dodonov, S.S. Mizrahi, A.L.S. Silva, Decoherence and thermalization dynamics of a quantum oscillator, 
J. Opt. B: Quantum Semiclass. Opt. 2 (2000) 271.
%
\bib{r37} J. Pleba\'{n}ski, Classical properties of oscillator wave packets, Bull. Acad. Pol. Sci. 2 (1954) 213.
%
\bib{r38} F.A.M. de Oliveira, P.L. Knight, V. Bu\v{z}ek, Properties of displaced number states, Phys. Rev. A 41 (1990)
2645.
%
\bib{r39} C. Monroe, D.M. Meekhof, B.E. King, S.R. Jefferts, W.M. Itano, D.J. Wineland, Resolved-Sideband Raman Cooling of
a Bound Atom to the 3D Zero-Point Energy, Phys. Rev. Lett. 75 (1995) 4011.
%
\bib{r40} C.A. Sackett, D. Kielpinski, B.E. King, C. Langer, V. Meyer, C.J. Myatt, M. Rowe, Q.A. Turchette, W.M. Itano, 
D.J. Wineland, C. Monroe, Experimental entanglement of four particles, Nature 404 (2000) 256.
%
\bib{r41} R.L. de Matos Filho, W. Vogel, Quantum Nondemolition Measurement of the Motional Energy of a Trapped Atom, Phys.
Rev. Lett. 76 (1996) 4520.
%
\bib{r42} D.J. Wineland, C. Monroe, W.M. Itano, D. Leibfried, B.E. King, D.M. Meekhof, Experimental Issues in Coherent
Quantum-State Manipulation of Trapped Atomic Ions, J. Res. Natl. Inst. Stand. Technol. 103 (1998) 259.
%
\bib{r43} D. Leibfried, R. Blatt, C. Monroe, D. Wineland, Quantum dynamics of single trapped ions, Rev. Mod. Phys. 75
(2003) 281.
%
\bib{r44} L. Davidovich, M. Orszag, N. Zagury, Quantum nondemolition measurements of vibrational populations in ionic 
traps, Phys. Rev. A 54 (1996) 5118.
%
\bib{r45} H.J. Carmichael, Statistical Methods in Quantum Optics 1: Master Equations and Fokker-Planck Equations, Springer, 
Berlin, 1999.
%
\bib{r46} S. Chountasis, A. Vourdas, Weyl functions and their use in the study of quantum interference, Phys. Rev. A 58 
(1998) 848.
%
\bib{r47} S. Chountasis, A. Vourdas, Weyl and Wigner functions in an extended phase-space formalism, Phys. Rev. A 58 (1998) 
1794.
%
\bib{r48} Recently, Turchette et al. \cite{r25} have investigated motional heating of laser-cooled 
$^{9}{\mbox {\rm Be}}^{+}$ ions held in radio-frequency (Paul) traps, and measured heating rates in a variety of traps with
different geometries, electrode materials, and characteristic sizes. This important experiment has shown that 
``the magnitude of heating rates of the size-scaling measurements are inconsistent with thermal electronic noise as the
source of the heating."
%
\bib{r49} A.I. Lvovsky, S.A. Babichev, Synthesis and tomographic characterization of the displaced Fock state of light,
Phys. Rev. A 66 (2002) 011801(R).
%
\bib{r50} G.N. Watson, A treatise on the theory of Bessel functions, Cambridge University Press, New York, 1996.
%
\bib{r51} N.N. Lebedev, Special Functions and Their Applications, Dover, New York, 1972.
%
\bib{r52} A.W. Niukkanen, Clebsch-Gordan-type linearisation relations for the products of Laguerre polynomials and
hydrogen-like functions, J. Phys. A: Math. Gen. 18 (1985) 1399.
%
\end{references}
\end{document}